\documentclass[fleqn,usenatbib]{mnras}

\usepackage{newtxtext,newtxmath}

\usepackage[T1]{fontenc}
\usepackage{ae,aecompl}

\usepackage{graphicx}	
\usepackage{amsmath}	
\usepackage{multicol}

\usepackage{xspace}
\usepackage{xifthen}

\newcommand{\roughly}{\ensuremath{ {\sim}\,} }

\newcommand{\code}[1]{\texttt{#1}\xspace}
\newcommand{\var}[1]{\ensuremath{\texttt{\MakeUppercase{#1}}}\xspace}

\newcommand{\unit}[1]{\ensuremath{\mathrm{\,#1}}\xspace}

\newcommand{\asec}{\unit{arcsec}}

\newcommand{\nm}{\unit{nm}}

\newcommand{\magn}{\unit{mag}}

\providecommand{\deg}{}
\renewcommand{\deg}{\unit{deg}}

\newcommand{\secref}[1]{Section~\ref{sec:#1}}
\newcommand{\appref}[1]{Appendix~\ref{app:#1}}
\newcommand{\tabref}[1]{Table~\ref{tab:#1}}

\newcommand{\figref}[1]{Figure~\ref{fig:#1}}
\newcommand{\figrefs}[2]{Figures~\ref{fig:#1} and \ref{fig:#2}}

\newcommand{\figrefss}[4]{Figures~\ref{fig:#1},~\ref{fig:#2},~\ref{fig:#3} and~\ref{fig:#4}}

\newcommand{\eqnref}[1]{Equation~\eqref{eqn:#1}}

\newcommand{\bandvar}[2][]{%
  \ifthenelse{\isempty{#1}}{\var{#2}}{\var{#2\_#1}}%
}

\newcommand{\nitermodel}[1][]{\bandvar[#1]{niter\_model}}

\newcommand{\sofmash}[1][]{\bandvar[#1]{extended\_class\_mash\_sof}}

\newcommand{\flagsfootprint}{\var{flags\_footprint}}

\newcommand{\flagsgold}{\var{flags\_gold}}
\newcommand{\flagssof}{\var{flags\_sof}}
\newcommand{\imaflags}{\var{imaflags\_iso}}

\newcommand{\ra}{{\ensuremath{\alpha_{2000}}}\xspace}
\newcommand{\dec}{{\ensuremath{\delta_{2000}}}\xspace}

\newcommand{\nz}{{\ensuremath{N(z)}}\xspace}

\newcommand{\photoz}{photo-$z$\xspace}
\newcommand{\pz}{\photoz}
\newcommand{\Photoz}{Photo-$z$\xspace}

\newcommand{\sof}{SOF\xspace}
\newcommand{\SOF}{\sof}
\newcommand{\ZMEAN}{\code{Z\_MEAN}}
\newcommand{\ZMC}{\code{Z\_MC}}

\newcommand{\SExtractor}{\code{SourceExtractor}}

\newcommand{\ngmix}{\code{ngmix}}

\newcommand{\iband}{$i$-band}

\newcommand{\redmagic}{redMaGiC}
\newcommand{\maglim}{MagLim}
\newcommand{\HEALPix}{\code{HEALPix}}
\newcommand{\healpix}{\HEALPix}

\newcommand{\nside}{\code{nside}}

\newcommand{\mangle}{\code{mangle}}

\newcommand{\bpz}{\code{BPZ}}
\newcommand{\dnf}{\code{DNF}}
\newcommand{\Annz}{\code{Annz2}}
\newcommand{\lephare}{\code{LePhare}}

\newcommand{\yonegold}{\code{Y1\,GOLD}}
\newcommand{\gold}{\code{Y3 GOLD}}

\newcommand{\baosample}{{\rm {BAO\,sample}}\xspace}

\usepackage{eso-pic}

\AddToShipoutPictureBG*{%
  \AtPageUpperLeft{%
    \hspace{0.75\paperwidth}%
    \raisebox{-3.5\baselineskip}{%
      \makebox[0pt][l]{\textnormal{DES-2019-0446}}
}}}%

\AddToShipoutPictureBG*{%
  \AtPageUpperLeft{%
    \hspace{0.75\paperwidth}%
    \raisebox{-4.5\baselineskip}{%
      \makebox[0pt][l]{\textnormal{FERMILAB-PUB-21-286-AE}}
}}}%

\title[DES Y3 \baosample]{Dark Energy Survey Year 3 Results: Galaxy Sample for BAO Measurement}

\author[A.~Carnero~Rosell et al.]{
\parbox{\textwidth}{
\Large
A.~Carnero~Rosell,$^{1,2,3}$
M.~Rodriguez-Monroy,$^{4}$
M.~Crocce,$^{5,6}$
J.~Elvin-Poole,$^{7,8}$
A.~Porredon,$^{7,8}$
I.~Ferrero,$^{9}$
J.~Mena-Fern\'andez,$^{4}$
R.~Cawthon,$^{10}$
J.~De~Vicente,$^{4}$
E.~Gaztanaga,$^{5,6}$
% ...
\begin{center} (DES Collaboration) \end{center}
}
\\
\\
Affiliations are listed at the end of the paper.
}

\date{Accepted XXX. Received YYY; in original form ZZZ}

\pubyear{2021}

\begin{document}
\label{firstpage}
\pagerange{\pageref{firstpage}--\pageref{lastpage}}
\maketitle

\begin{abstract}

In this paper we present and validate the galaxy sample used for the analysis of the Baryon Acoustic Oscillation signal (BAO) in the Dark Energy Survey (DES) Y3 data. The definition is based on a colour and redshift-dependent magnitude cut optimized to select galaxies at redshifts higher than 0.5, while ensuring a high quality determination. The sample covers $\roughly 4100$ square degrees to a depth of $i = 22.3 \ (AB)$ at $10\sigma$. It contains 7,031,993 galaxies in the redshift range from $z$= 0.6 to 1.1, with a mean effective redshift of 0.835. Redshifts are estimated with the machine learning algorithm \dnf, and are validated using the VIPERS PDR2 sample. We find a mean redshift bias of $z_{\mathrm{bias}} \roughly 0.01$ and a mean uncertainty, in units of $1+z$, of $\sigma_{68} \roughly 0.03$. We evaluate the galaxy population of the sample, showing it is mostly built upon Elliptical to Sbc types. Furthermore, we find a low level of stellar contamination of $\lesssim 4\%$. We present the method used to mitigate the effect of spurious clustering coming from observing conditions and other large-scale systematics. We apply it to the \baosample and calculate weights that are used to get a robust estimate of the galaxy clustering signal. This paper is one of a series dedicated to the analysis of the BAO signal in DES Y3. In the companion papers we present the galaxy mock catalogues used to calibrate the analysis and the angular diameter distance constraints obtained through the fitting to the BAO scale.
\end{abstract}

\begin{keywords}
cosmology: observations - (cosmology:) large-scale structure of Universe - catalogues - surveys
\end{keywords}



\section{Introduction}

Baryon Acoustic Oscillations (BAO) is one of the most remarkable predictions of the formation of structures in the Universe~\citep{Peebles1970, SZ1970, Bond1984, Bond1987}. Since its confirmation in the distribution of galaxies in 2005~\citep{Eisenstein2005}, BAO measurement has been one of the main scientific drivers in the design and construction of galaxy surveys. 

BAO has already been detected several times in spectroscopic~\citep{Cole:2005, Percival:2007, Gaztanaga:2009, Percival:2010, Beutler:2011, Blake:2011, Ross:2015, Alam2017} and photometric~\citep{Padmanabhan:2007, Estrada:2009, Hutsi:2010, Crocce:2011, carnero2012, y1bao} data-sets, for galaxies, but also in the distribution of QSOs~\citep{Ata2018} and Lyman-alpha absorbers~\citep{Bautista2017}, in a wide variety of redshifts, from $z=0.2$ to $z<3$. Estimates of the evolution of the BAO scale with time is a direct measurement of the expansion of the Universe and therefore, an excellent cosmology observable. All these measurements are compatible with the $\Lambda$CDM cosmological model and the existence of Dark Energy. 

In this context, the Dark Energy Survey (DES:~\citealt{Flaugher:2005, DES:2016ktf}) set as one of its main objectives to measure the BAO scale in the distribution of galaxies. In a previous release, in DES Year 1~\citep{y1bao} we measured the BAO scale at an effective redshift of 0.81. This sample covered approx. 1400 square degrees; given the limited area, this detection had a low significance. In DES Year 3 (Y3), the data-set analysed here, the nominal footprint of approx. 5000 square degrees for the complete survey is covered, and we expect to reach a sensitivity to BAO of the same order as concurrent spectroscopic and photometric surveys. Furthermore, this measurement will be combined with the other DES cosmological observables to estimate the most precise measurements on Dark Energy by combination of BAO with Weak Lensing, Supernova Ia and Galaxy Clusters number counts. 

One of the main difficulties in detecting BAO in photometric surveys is the smearing in the signal produced by the poor redshift determination. In this context, it is necessary to select a galaxy population that presents a prominent spectral feature that can be captured with broadband filters. Generally, the best practice is to select old, well-evolved galaxies with a significant Balmer break~\citep{Eisenstein:2001, Vakili:2019, baosample, Zhou:2020}. This feature makes galaxies look very red, and they usually drive the target selection in galaxy surveys. 

In~\citet{baosample}, we developed a colour selection to select galaxies in the DES Y1 sample, calibrated through a set of synthetic SED distributions, optimized for redshifts $z>0.5$. In this new release, we apply the same colour selection, but put the focus in the improvement of ameliorating spurious clustering due to observing conditions. Since the sample now covers more than 4000 square degrees, with observations taken during three different years, variations on conditions like seeing, airmass, sky brightness, stellar density, or galactic extinction are expected to leave significant imprints in the galaxy clustering and therefore, robust corrections are needed. In DES Y3 we compute weights to correct for these effects, following the procedure developed in~\citet{elvinpoole} for the DES Y1 lens sample, but to the DES Y3 \baosample. Similar methods have also been applied to BOSS~\citep{ross11, 2017MNRAS.464.1168R}, eBOSS~\citep{2020MNRAS.498.2354R, Laurent_2017} and DESI~\citep{10.1093/mnras/staa1621} targets.

The structure of the paper is as follows: In \secref{y3gold} we present the parent DES Y3 data and next, in~\secref{baosample} we describe the \baosample selection. In \secref{footprint} we present the selection of the footprint. In \secref{photoz} we present the photometric redshift (\pz) definition and validation. In \secref{valid} we validate the sample. In \secref{blinding} we summarize the blinding procedure, which is largely shared with other DES analyses. In \secref{lsssys} we present the analysis of mitigation of observational systematic effects and in~\secref{unblinding} we show the 2-pt clustering signal in real space after the unblinding of the sample. Finally, we show our conclusions in~\secref{conclusions}. The \baosample will be eventually released at \url{https://des.ncsa.illinois.edu/releases} together with all the DES Y3 products.  

This paper accompanies a series of papers focused on the analysis of BAO distance in DES Y3 analysis. In~\citet{colamocks} we construct precise galaxy mocks for the \baosample, that are used to validate and optimise the analysis. Finally, in~\citet{y3mainbao} we measure the BAO scale as a function of redshift, both in real and in harmonic space and determine the best fit cosmological parameters for the \baosample.

It is worth noting that the study of the largest scales in galaxy clustering (through the angular correlation function and the angular power spectrum), not only allows the determination of the BAO scale, but the same observables can be used to study, for example, primordial non-gaussianities and neutrino mass, which will be the focus of future analyses.


\section{DES Y3 Data}
\label{sec:y3gold}

The Dark Energy Survey operations ended in 2019, after six years. DES used the Blanco 4m telescope at Cerro Tololo Inter-American Observatory (CTIO) in Chile, and  observed $\roughly 5000 \deg^2$ of the southern sky in five broadband filters, $grizY$, ranging from $\roughly400 \nm$ to $\roughly1060 \nm$ \citep{Li:2016,Y3FGCM}, using the DECam~\citep{Flaugher:2015} camera. Finally, images are processed at DES Data Management in NCSA~\citep{desdm}.

The \baosample was selected based on the \gold catalogue~\citep{y3gold}, an improved version of the public DR1 data~\citep{dr1}\footnote{Available at \url{https://des.ncsa.illinois.edu/releases/dr1}}. It includes observations from the first three years of operations. In comparison with the DR1 public release, \gold includes improved photometric zero-point corrections, several observing condition maps, more advanced photometry extraction, morphological star/galaxy separation, and extra quality flags not included in DR1. For details in the construction of the \gold catalogue, we refer to \citet{y3gold}. In this section, we briefly detail those quantities that are of importance to the selection of the \baosample.

Photometric information is obtained through the multi-epoch, multi-band fit to objects based on the \ngmix software~\citep{y1gold}. In DES Y3, we run in a simplified mode that eliminates the multi-object light subtraction step. This speeds the process as well as ensures fewer objects have fit failures. This mode is called \SOF (single-object-fitting), and it is the base for the \baosample. So far, \SOF photometry was only calculated for $griz$ but not the $Y$ band. Internal studies showed that the $Y$ band's use did not improve redshift estimations in Y3 and therefore the band is not used for the \baosample. Furthermore, using \SOF morphological information, a star/galaxy classification method is developed, by grouping objects according to their similarity with a point-like or extended source. 

Several zero-point corrections are applied to the grey calibration presented in DR1: additional corrections based on observations from up to Y4 observations and chromatic corrections, which are spectral energy distribution (SED) dependent zero-point corrections to correct for differences to the standard star used in calibration~\citep{Li_2016}. Furthermore, galactic reddening is also corrected as a function of the SED of the source.

\Photoz's are calculated using $griz$ \SOF corrected fluxes, trained on a large spectroscopic sample that includes public as well as private spectral references~\citep{Gschwend:2018}. During the creation of the \baosample, we tried other photometric redshift estimates using \SExtractor \texttt{AUTO} fluxes for \pz, as well as using different \pz codes, but \SOF based \dnf gave the best metrics in all cases, using an independent validation sample.

Finally, depth maps were also built for galaxies for \SOF corrected magnitudes, used along with several observing conditions maps to select the effective area of DES Y3 analysis and for mitigation of spurious clustering due to observing conditions. 


\subsection{Changes with respect to DES Y1}

Several changes have been applied to the DES Y3 data processing pipeline improving over Y1 reduction. This leads to a greater number density of galaxies in Y3 compared to Y1, and allows the extension of the \baosample to $z>1.0$. Even though the mean number of exposures in each position is similar to Y1 (around four exposures in $griz$), the homogeneity is larger for Y3, resulting in a slight increase in depth ($\sim 0.02\magn$ in the bluest bands to $\sim 0.2\magn$ in the reddest at $snr=10$). Also, modifications in \SExtractor settings have altered the number of objects detected, reaching lower signal-to-noise than in Y1. 

\begin{figure}
    \includegraphics[width=\linewidth]{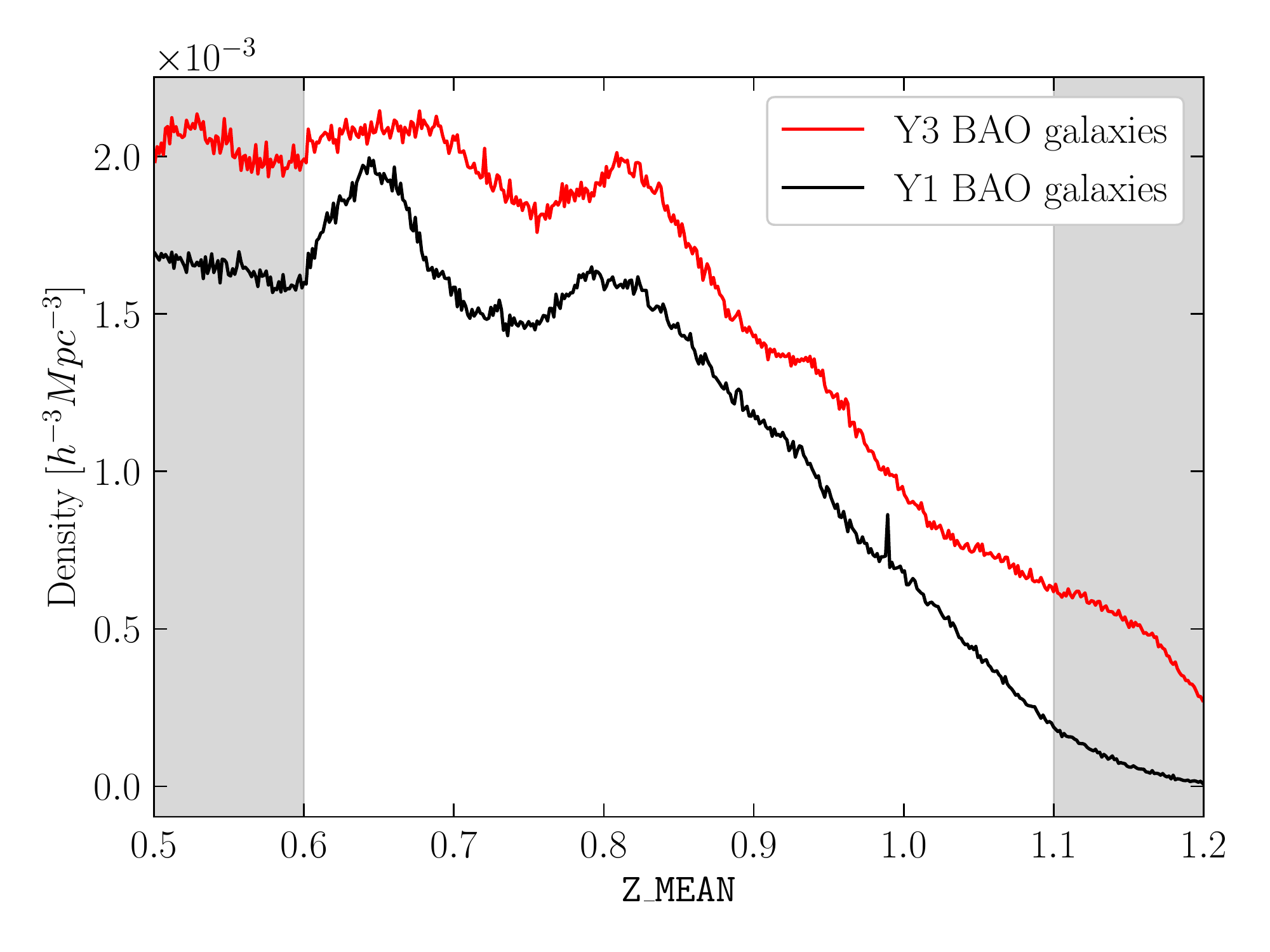}
\caption{Number density of galaxies found in \yonegold (black) and \gold (red). The increase is mostly due to changes in the data reduction process. This increased justified the extension of the galaxy sample to redshift $z=1.1$. 
}
    \label{fig:y1y3_numberdens}
\end{figure}

As previously stated, these changes make Y3 galaxy samples denser, as can be seen in \figref{y1y3_numberdens}, with an increase of a factor $\approx 1.3$ in galaxies per $h^{3}Mpc^{-3}$.

\section{Sample selection}
\label{sec:baosample} 

The \baosample has been selected following the same procedure as that of DES Y1 BAO analysis. We first apply the following quality cuts to \gold data~\citep{y3gold}:

\begin{itemize}
    \item \flagsgold: we remove, for all bands $griz$, any source with $\flagssof \ != 0$, with \SExtractor $\var{flags} > 3$, or $\imaflags \ != 0$. As well as any object defined as \emph{Bright Blue artifact} or \emph{Bright objects with nonphysical colors and possible transients}.
    
    \item \flagsfootprint. We require sources to be defined inside the footprint and to have $\nitermodel > 0$ in all $griz$ bands to ensure that the object has been observed. The footprint is defined as those regions with at least one exposure in all $griz$ bands, with a coincident effective coverage greater than 50\% in all bands. 
    
    \item We further impose sources to be within the angular mask, defined in \secref{footprint}.
    
\end{itemize}

Once we apply the quality cuts defined above, we select secure galaxies using the \sofmash classifier, which classifies how much a source differs from a point-like morphology~\citep{y3gold}. For \pz estimates (see \secref{photoz}) and for the galaxy colour selection, we use \SOF magnitudes corrected by SED-dependent extinction based on SFD98~\citep{sfd98}, chromatic corrections and \gold zero-point corrections. 

We next apply the colour selection:

\begin{equation}
\label{eqn:colorcut}
 (i-z) + 2\times (r-i) > 1.7 \ ,
\end{equation}

\noindent and also the flux-limit cut:

\begin{equation}
\label{eqn:fluxlimit}
 i < 19. + (3 \times \ZMEAN) \ ,
\end{equation}

\noindent in the magnitude range $17.5<i<22.3$. Finally we select galaxies in the redshift range:
\begin{equation}
\label{eqn:redshiftlimit}
 0.6 < \ZMEAN < 1.1 \ .
\end{equation}

The flux-limit cut, in the presence of large \pz scatter, could lead to unwanted correlations with other survey conditions and to an amplification of systematic effects. In our case, the \baosample is, by definition, designed to avoid large \pz scatters and therefore, these correlations will be small. Likewise, this effect should be corrected by the amelioration of survey conditions correlations described in~\secref{lsssys}.

Comparing our selection algorithm with the one used for the DES Y1 (see equation 3 in \citealt{baosample}), we have increased the depth limit up to 22.3\magn to reach redshifts $z \leq 1.1$ after considering Equation~\ref{eqn:fluxlimit}. This, in turn, implies a reduction of an area of 100 square degrees, in comparison with a sample with a depth limit of 22\magn, reaching redshifts up to $z=1$. Based on cosmological forecasts, adding this extra bin at $z<1.1$ at the expense of losing 100 square degrees results in better constraining power. We predict a gain of 10\% in the precision of the BAO scale and a higher mean redshift.

\begin{table*}
\centering
\caption{Summary of the selection process. Starting from the \gold catalogue, we apply the same colour selection as in \yonegold, but we extend the analysis to redshift $z=1.1$. We also need to extend the magnitude limit to $i<22.3$, ensuring completeness through the entire footprint.}
\label{tab:bao_selection}
\begin{tabular}{c c c}
\hline
\hline
Keyword & Cut &  Description \\
\hline
Gold & observations present in the \gold catalogue & \citet{y3gold} \\
Quality & \flagsgold & \secref{baosample} \\
Footprint & 4108.47 $deg^{2}$ & \secref{footprint} \\
Colour selection & $(i-z) + 2\times (r-i) > 1.7$ & \secref{baosample} \\
Flux Selection & $17.5< i<19. + 3. \times \ZMEAN$ & \secref{baosample} \\ 
Star-galaxy separation & $\sofmash=3$ & \secref{baosample} \\
\Photoz range & [$0.6-1.1$] & \secref{photoz} 
 \\
\end{tabular}
\end{table*}

In \tabref{bao_selection}, we list all the selections done in the \gold sample to select the \baosample. In \tabref{bao_properties} we summarize the main properties of the \baosample for each tomographic redshift bin. Redshift properties are explained in \secref{photoz}. The galaxy bias estimates of this table have been obtained following the blinding procedure explained in \secref{blinding} and serve as initial values for subsequent analyses. 

\begin{table*}
\centering
\caption{Main properties of the \baosample in each tomographic bin. Redshift properties are given for the VIPERS sample estimation, namely, the mean redshift ($\bar{z}$), the width of the \nz ($W_{68}$) and the dispersion on the \pz error ($\sigma_{68}$). Sample variance are estimated based on MICE simulation. The so-called blind galaxy bias has been obtained following the collaboration's blinding procedure and serve as initial values needed to mitigate systematic observational effects.}
\label{tab:bao_properties}
\begin{tabular}{c c c c  c c}
Redshift limits & $\bar{z}$ &  $W_{68}$ & $\sigma_{68}$ &  Number of galaxies & blind galaxy bias \\
\hline
0.6 < z < 0.7 & 0.648 $\pm$ 0.003  & 0.0455 $\pm$ 0.003  & 0.021 $\pm$ 0.001 &   1,478,178 & 1.79 $\pm$ 0.09 \\
0.7 < z < 0.8 & 0.742 $\pm$ 0.003  & 0.0522 $\pm$ 0.002  & 0.025 $\pm$ 0.002 &   1,632,805 & 1.83 $\pm$ 0.10 \\
0.8 < z < 0.9 & 0.843 $\pm$ 0.003  & 0.0629 $\pm$ 0.003  & 0.029 $\pm$ 0.002 &  1,727,646 & 2.02 $\pm$ 0.12  \\
0.9 < z < 1.0 & 0.932 $\pm$ 0.004  & 0.0633 $\pm$ 0.003  & 0.030 $\pm$ 0.003 &  1,315,604 & 2.09 $\pm$ 0.14 \\
1.0 < z < 1.1 & 1.020  $\pm$ 0.006  & 0.0808 $\pm$ 0.006  & 0.040 $\pm$ 0.005 &  877,760 &  2.4 $\pm$ 0.08  \\
\end{tabular}
\end{table*}

\section{Angular Mask}
\label{sec:footprint}

Apart from the object-to-object selection of the \baosample, we need to define the effective area of the sample, ensuring we remove areas of dubious quality and that are complete given the magnitude limit of the \baosample. In DES, the exact image footprint information is delivered as \mangle products~\citep{mangle}, which are later transformed into \healpix maps~\citep{Gorski:2005} of resolution \nside=4096. This translation facilitates merging several information maps, which can then be combined to generate the sample's footprint.

The \baosample is only defined in the Wide Survey area, excluding the Supernova Fields (see~\citealt{y3gold} for details). To create the footprint mask, we impose the following requirements in the input \healpix maps. For details about the maps we refer to \citet{y3gold}:
\begin{itemize}
    \item At least one exposure in each of the $griz$ bands.
    \item The effective area of each pixel must be greater than 0.8 in $griz$. This cut removes \healpix pixels lying at the edge of the survey area, or containing significantly masked area.
    \item Pixels must not be affected by foreground sources, like regions around bright stars or extended galaxies.
    \item An extra cleaning is done in very obvious pixels affected by scattered light and unmasked streaks.
    \item Pixel must have a \SOF corrected $10\sigma$ depth in $griz$ greater than 22, 22, 22.3, 21\magn, respectively.
    \item The depths in $r$ and $z$ bands must follow the condition, $(2 \times r_{\mathrm{lim}}-z_{\mathrm{lim}})>24$. This condition is to ensure that our sample selection in~\eqnref{colorcut} remains complete and applicable following the minimum depth cut in $i$-band ($i>22.3$).

\end{itemize}

\begin{figure}
    \includegraphics[width=\linewidth]{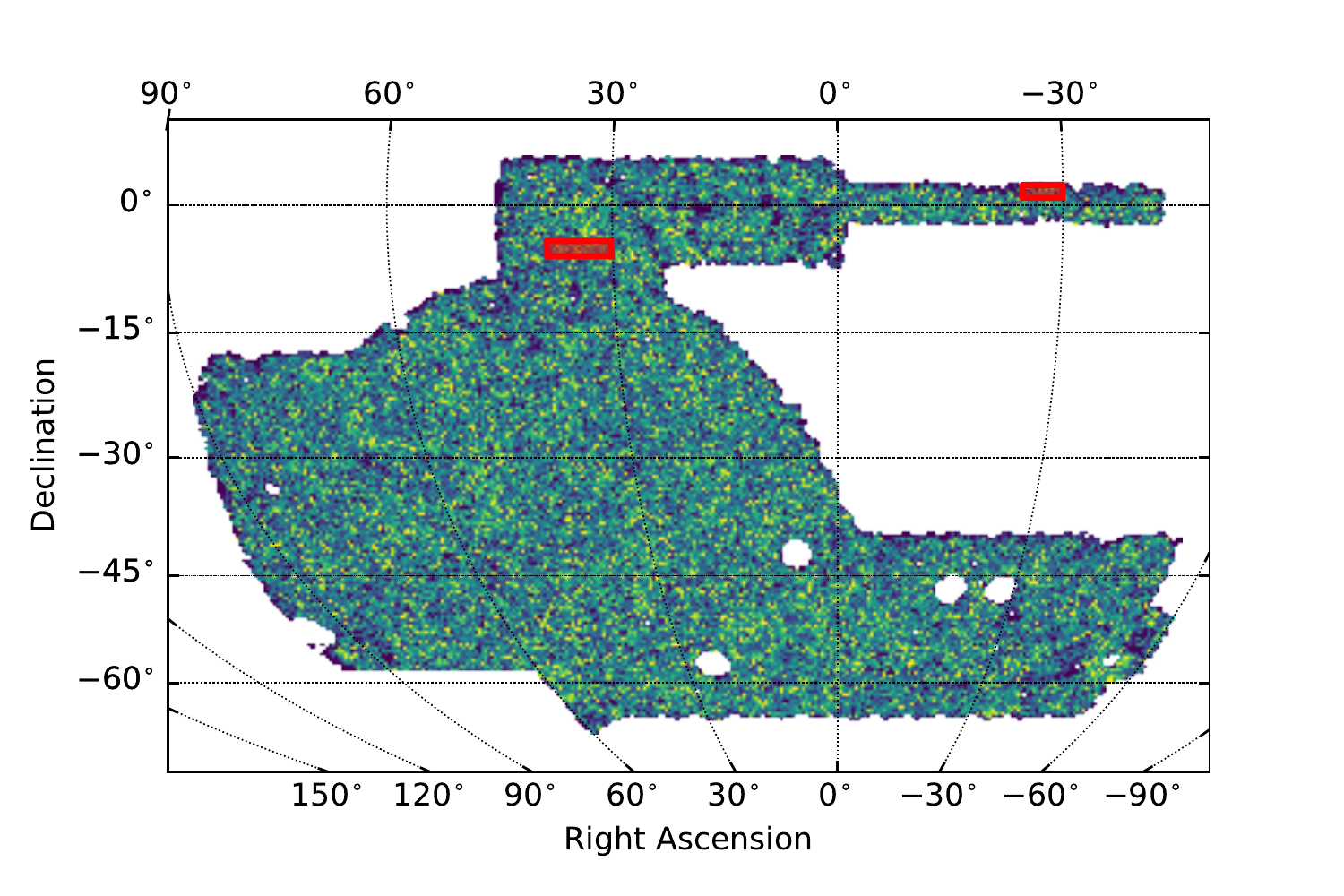}
\caption{\baosample angular distribution, covering 4108.47 square degrees of the sky accessible at CTIO. In red squares, VIPERS W1 and W4 regions, used for \pz validation.}
    \label{fig:footprint}
\end{figure}

After combining all these conditions, we end up with the final \baosample mask, which covers 4108.47 square degrees. The footprint and density distribution can be seen in \figref{footprint}. Here, the effective area of the pixels are considered. In this figure we also show in red the W1 and W4 VIPERS regions, used for \pz calibration in \secref{vipers}. 

\section{Photometric redshifts}
\label{sec:photoz}

\Photoz's used for the \baosample are derived using the Directional Neighborhood Fitting (\dnf) algorithm \citep{dnf}. We use the \pz estimates described in \gold~\citep{y3gold}. In summary, \dnf was trained using the most updated spectroscopic sample available for the collaboration by May 2018. This sample contains public spectroscopic but also some DES proprietary spectroscopic data, including the collection from the OzDES collaboration~\citep{ozdes20}. From this sample, we randomly selected half of the objects for training, while the rest was left for general validation purposes. Furthermore, we also removed the whole VIPERS spectroscopic sample from the training sample, and we left it for validation purposes, as explained in the following subsection. To estimate \pz's, we used \SOF corrected fluxes, using $griz$ bands.

In internal analyses, we also tested \bpz code \citep{Benitez:2000} and \Annz \citep{Sadeh:2015}, but \dnf always gave better \pz bias and $\sigma_{68}$ metrics, and therefore, we use it to estimate redshifts for our \baosample.

In addition to the predicted best-value in the fitted hyper-plane (\ZMEAN), \dnf also returns the redshift of the nearest neighbour (\ZMC). This quantity, stacked for all galaxies in a given tomographic bin, has proven to give a fair description of the true \nz~\citep{baosample}. Likewise, \dnf produces redshift probability distributions (\var{PDF}) for each source, that can also be used to estimate the \nz of a given tomographic bin if we stack the \var{PDF} of all the selected sources. Both estimates will be used to validate our fiducial \nz, based on the VIPERS spectroscopic survey. 




To estimate the \nz of the \baosample we use the second public data release \citep[PDR2]{vipers_pdr2} from the ``VIMOS Public Extragalactic Redshift Survey'' \citep[VIPERS]{vipers}. Unlike with the DES Y1 analysis, where we used the COSMOS sample~\citep{Laigle:2016}, we employ VIPERS, which is a larger sample and, therefore, less affected by cosmic variance. Likewise, VIPERS is complete up to $i_{AB} \approx 22.5$ for redshifts above 0.5, where the \baosample is defined.

\subsection{VIPERS Validation}
\label{sec:vipers}


The VIPERS sample consists of 91,507 sources, from which 86,775 are galaxies. VIPERS observed in two fields, named W1 and W4, both overlapping DES. W1 is the largest area, centred at $\ra = 34.495^{\circ}$ and $\dec = -5.076^{\circ}$, and covers an effective area of 11.012 square degrees, while W4 is centred at $\ra = 332.7^{\circ}$ and $\dec = 1.61^{\circ}$, and covers 5.312 square degrees. The total overlap area is 16.324 square degrees (seen in \figref{footprint}).

The sample was defined to be statistically complete above redshift 0.5, at least up to $i_{AB}<22.5$~\citep{vipers_pdr2}. Considering that the \baosample is defined above redshift 0.6 and up to $i_{AB}<22.3$ makes VIPERS an excellent reference sample for the \pz validation.

As recommended by ~\citet{vipers_pdr2}, we apply the following quality cuts on the VIPERS data\footnote{\url{http://vipers.inaf.it/data/pdr2/catalogs/PDR2_SPECTRO_TABLES.html}}: 
\begin{itemize}
    \item $2 \leq zflag < 10$: ensures a good quality spectroscopic redshift with more than 90\% confidence and eliminates AGNs and duplicated objects.
    
    \item The Target Sampling Rate ($tsr$), the Spectroscopic Success Rate ($ssr$) and the Colour Sampling Rate ($csr$) must be greater than zero. 
    
    \item $classFlag==1$: selects from the main catalog, galaxies with colours compatible with $z>0.5$.
    
    \item $photoMask==1$ ensures that galaxies fall within the photometric mask. 
\end{itemize}

These quality cuts are more than sufficient for our analysis, since cosmological constraints coming from the BAO scale measurement are robust against a small number of wrong redshifts in the determination of the \nz.

After applying these cuts, we end up with 74,591 galaxies available for \pz validation and estimation of our \nz's. Furthermore, VIPERS statistics must be weighted to account for the various sampling rates. The galaxy weight for each VIPERS galaxy is $w=1./ssr/tsr/csr$.

We then match the VIPERS sample to the \baosample in separated redshift bins. In total, we have 8362 galaxies matched within 1 \asec, divided into tomographic redshift bins the number of galaxies available for calibration is $1934,~ 2107,~ 2167,~ 1416,~ 738$ respectively. This represents $\roughly 12\%$ of the VIPERS galaxy sample. The reduction of the sample comes mostly from the colour selection and the flux selection which by themselves eliminates 82\% of the VIPERS catalogue. Despite this reduction, the mean value of $ssr \times tsr \times csr$ in each redshift bin is 0.47, 0.47, 0.45, 0.42, 0.39 respectively, which reflect the completeness of the validation sample with respect to the complete VIPERS sample. We further confirmed the VIPERS sample covers the same colour-colour space as the \baosample (cf.~\appref{mice_phot}).

We estimate the \baosample \xspace \nz in each tomographic bin by stacking the VIPERS redshifts for all the matches in the given redshift bin, but before, we validate \dnf \pz point estimates (used to assign galaxies to each tomographic bin) using the VIPERS redshifts (see \secref{photoz_valid}). The variance in the validation metrics and in the \nz's are obtained based on $\approx 316$ \baosample \pz realizations using the MICE simulation~\citep{mice, mice15}, as described in the following subsection. 

\subsection{Photo-$z$ variance}
\label{sec:nz_samplevar}


To estimate the uncertainties of the \pz metrics and of the \nz's, we rely on the MICE simulation to create several realizations of the VIPERS-\baosample. The MICE simulation covers $~\roughly 5150$ square degrees. Based on the method of~\citet{Lima:2008}, we create a MICE catalogue with the same photometric properties as the \gold. From here, we divide the MICE simulation in a total of 316 equal-area sub-samples of 16.3 square degrees, each consisting of two independent regions of 11 and 5.3 square degrees. From here, we select galaxies with the same cuts as in the \baosample. In each realization, we calculate the \nz ($316$ \nz for each redshift bin). Since MICE's galaxy density is a little higher than VIPERS density, in each of these sub-samples, we randomly pick as many objects as the VIPERS catalogue has, independently for each tomographic bin. 

In \appref{mice_phot}, we show a comparison of the properties of the \baosample and MICE simulation, confirming the simulation reproduce the photometry of the VIPERS-\baosample correctly (\figrefs{hist_mag}{colorcolor_diagrams}). Hence, we can use MICE to estimate errors of the \pz metrics and of the \nz's, as shown in~\figrefss{photoz_mean_vs_z_comparison}{photoz_sig68_vs_z_comparison}{nz_1}{photoz_W68_vs_z_comparison}.


\subsection{Photo-$z$ validation}
\label{sec:photoz_valid}

We assess the quality of the \dnf point \pz estimates by measuring the \pz bias and $\sigma_{68}$ in each tomographic bin. These two quantities are the most important ones to estimate the correct angular distance to the BAO. The outlier fraction, a standard metric given when assessing \pz's in generic studies, does have a negligible effect on BAO measurements, for example. 

We then calculate these two metrics for our \pz choice, based on VIPERS, and also for those estimates based on \ZMC. \ZMC is estimated for each \baosample source according to its nearest neighbour in the training sample. But the training sample is not constructed to represent the \baosample and therefore, their metrics are expected to be not as representatives as the VIPERS estimates.

The \pz bias is defined, averaged over all the available galaxies $N$, as:
\begin{equation}
    z_{\mathrm{bias}} = \frac{1}{N}\sum_{i=1}^{N}(\ZMEAN^{i}-z^{i}_{\mathrm{spec}})
\end{equation}

\noindent and $\sigma_{68}$ is defined as the value such as 68\% of the galaxies have $\lvert \ZMEAN-z_{\mathrm{spec}} \rvert / (1+z_{\mathrm{spec}}) < \sigma_{68}$.

\begin{figure}
    \includegraphics[width=\linewidth]{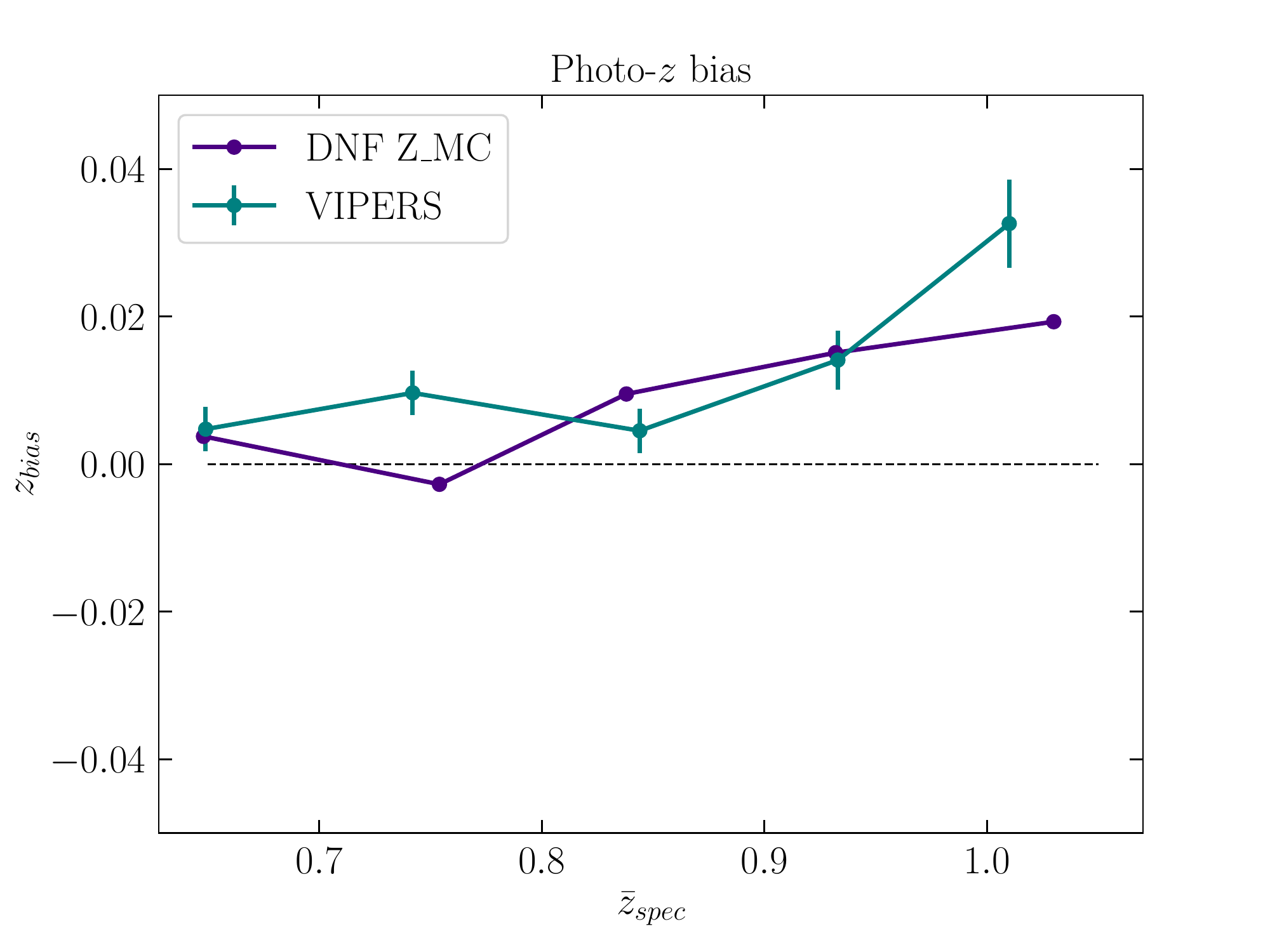}
\caption{$z_{\mathrm{bias}}$ as a function of $\bar{z}_{\mathrm{spec}}$, for the two $z_{\mathrm{spec}}$ estimates (VIPERS and \ZMC). $z_{\mathrm{bias}}$ is defined as the average difference between \ZMEAN and the given redshift. After finding a good agreement between independent estimates, we assume the VIPERS validation as our default choice. Errors only apply to the VIPERS sample, and they are estimated as the standard deviation from realizations (see \secref{nz_samplevar}).} 
    \label{fig:photoz_mean_vs_z_comparison}
\end{figure}

In \figrefs{photoz_mean_vs_z_comparison}{photoz_sig68_vs_z_comparison} we show the evolution of $z_{\mathrm{bias}}$ and $\sigma_{68}$ as a function of $\bar{z}_{\mathrm{spec}}$. Error bars are the sample variance calculated as the standard deviation from the MICE realizations. The level of $\sigma_{68}$ is below 0.1 and the mean bias is below 0.04 in all the redshift ranges, confirming the good quality of the \pz estimate, as expected by design of the sample selection.

\begin{figure}
    \includegraphics[width=\linewidth]{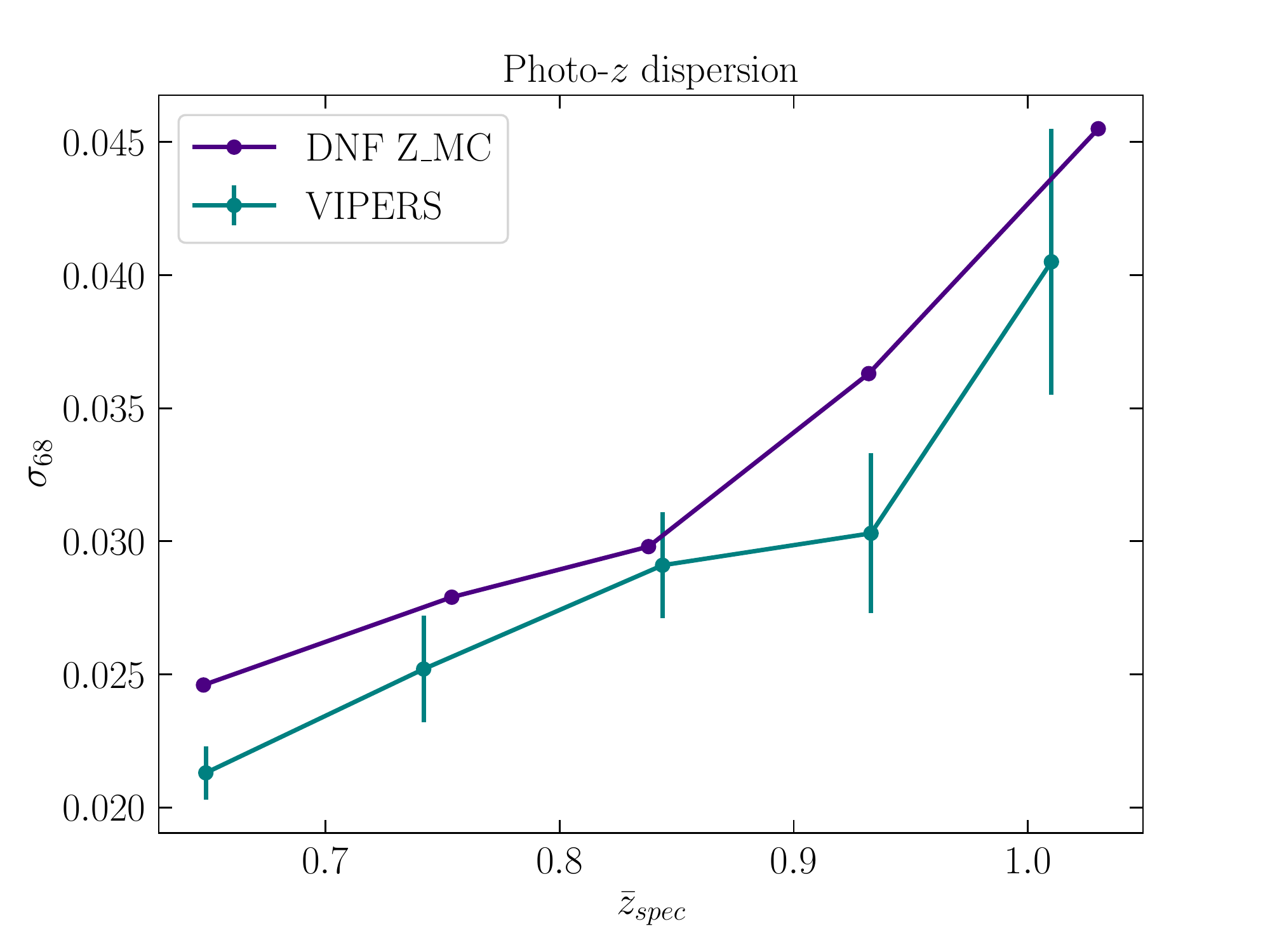}
\caption{Redshift bias scatter for the two redshift estimates (VIPERS and \ZMC) as a function of $\bar{z}_{\mathrm{spec}}$. $\sigma_{68}$ is defined as the value where 68\% of the objects have $\lvert \ZMEAN-z_{\mathrm{spec}} \rvert / (1+z_{\mathrm{spec}}) < \sigma_{68}$.}
    \label{fig:photoz_sig68_vs_z_comparison}
\end{figure}

\begin{figure*}
    \includegraphics[width=\linewidth]{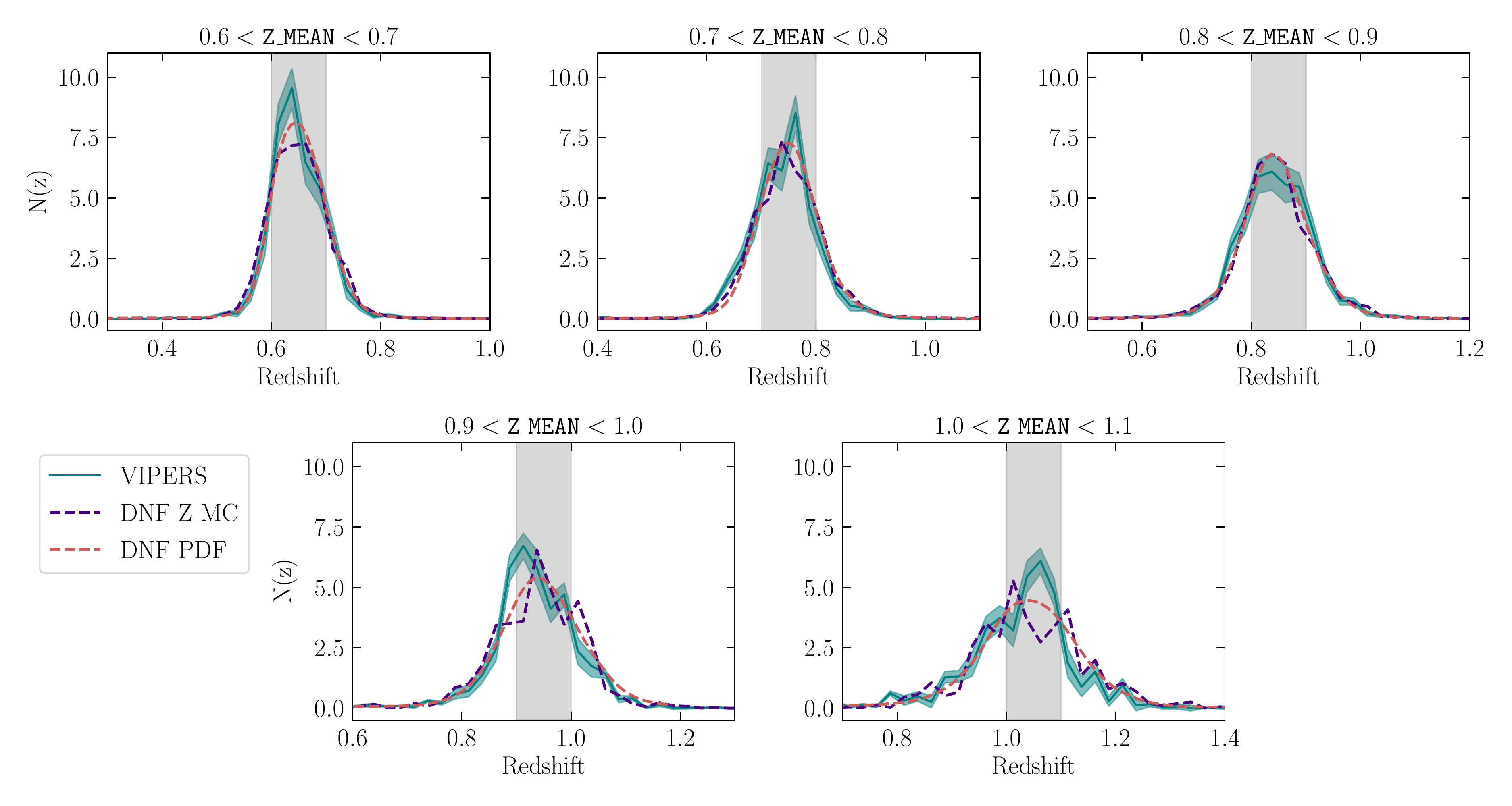}
\caption{Redshift distributions, \nz, estimated using three different data-sets. We first use DNF \ZMEAN to select galaxies in tomographic bins and then, we estimate the true \nz using the VIPERS sample. Errors are estimated as the standard deviation from 316 realizations (see \secref{nz_samplevar}). The \ZMC and \dnf \var{PDF} estimates are only shown for comparison.}
    \label{fig:nz_1}
\end{figure*}

\begin{figure}
    \includegraphics[width=\linewidth]{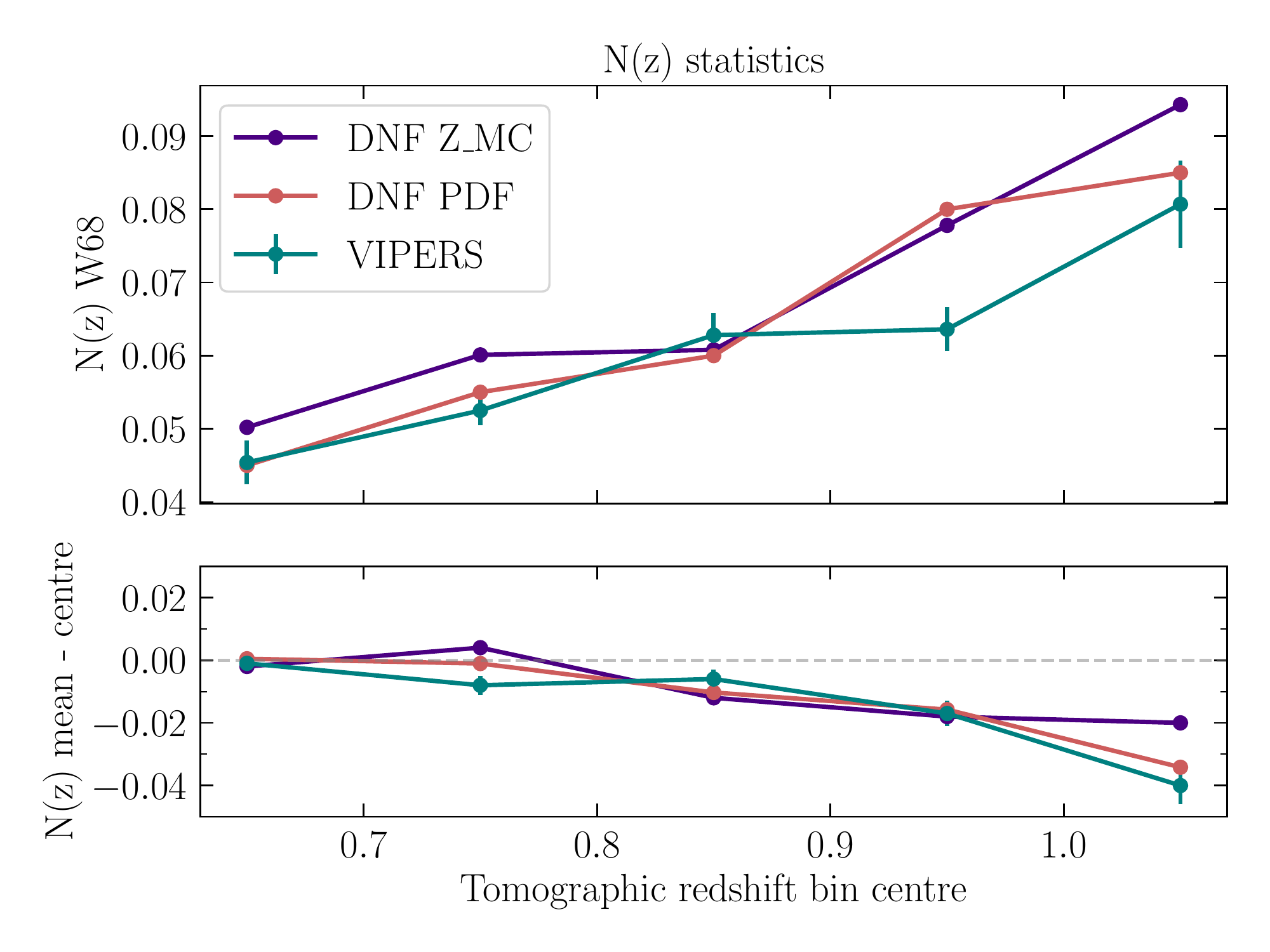}
\caption{Summary statistics for the \nz distributions shown in~\figref{nz_1}. On the top panel, the width of the distribution, $W_{68}$, as a function of the centre of the tomographic redshift bin. $W_{68}$ defines the width containing 68\% of the \nz distribution. On the bottom panel, the difference between the mean of the \nz and the centre of the bin. Errors only apply to the VIPERS sample, and they are estimated as the standard deviation from 316 mock realizations from MICE (see \secref{nz_samplevar}).}
    \label{fig:photoz_W68_vs_z_comparison}
\end{figure}

Once we confirm the quality of the \dnf \pz estimates, we proceed to estimate the \emph{true} \nz's in each bin. In \figref{nz_1} we present the \nz for each tomographic bin estimated with VIPERS, \dnf \ZMC and \dnf \var{PDF}. The variance is obtained based on the standard deviation estimated with MICE simulation. Finally, in \figref{photoz_W68_vs_z_comparison}, we present $W_{68}$, defined as the width containing 68\% of the \nz distribution, and the mean of \nz as a function of redshift for each calibration sample.


In general, \dnf behaves well, confirming \ZMC is a good tracer of the VIPERS redshifts. As already said, several internal tests showed that the choice of one or another \nz does not change our results.

\section{Sample validation}
\label{sec:valid}

In this section we validate the photometric properties of the \baosample and estimate the purity of the sample. 


\subsection{Photometry}
\label{sec:photometry}

The colour selection for the \baosample was defined to select galaxies beyond $z=0.5$, following the SED for elliptical galaxies. Details about the selection can be found in Figure 5 from~\citet{baosample}. There we estimated the colours of a set of SED templates as a function of redshift seen through the DES filter passbands. To confirm that the colour selection is still valid for Y3, we run the \baosample through the SED template fitting code \lephare~\citep{Arnouts:2011}, with the same SED templates as in Y1~\citep{Benitez:2000} and using the most updated versions of the DES filters passbands\footnote{\url{http://www.ctio.noao.edu/noao/content/DECam-filter-information}}. 

\begin{figure}
    \includegraphics[width=\linewidth]{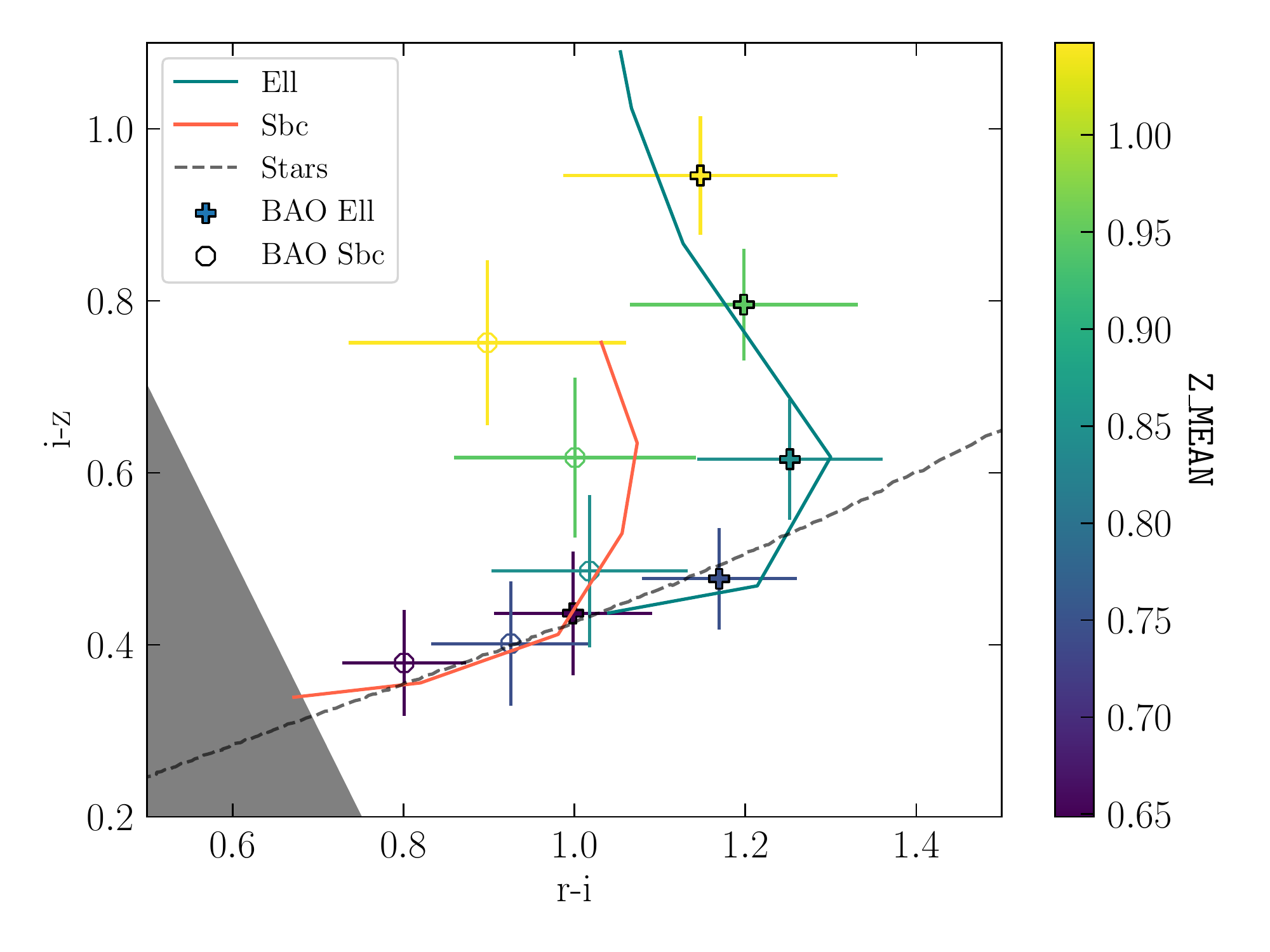}
\caption{Median colour-colour distributions for the \baosample, divided into spectral types, as a function of \ZMEAN redshift, compared with the expected colour-colour evolution of Elliptical and Sbc SEDs (solid lines). In dashed line, the \gold stellar locus. The white area defines the colour-colour cut applied to select the \baosample. Error bars are the colour standard deviations of each sub-sample.}
    \label{fig:colorcomp}
\end{figure}

We run \lephare with fixed redshifts to \ZMEAN (we also tested allowing redshifts to vary freely, and results were the same). With the best fitting model for each galaxy, we can examine the proportions of the different spectral type populations within the \baosample. After running \lephare, we find that 44\% of the sample are actually elliptical galaxies, 34\% Sbc types, and 22\% other types of galaxies. Furthermore, the colours reproduce well the expected locus for each spectral type. This can be seen in~\figref{colorcomp}, where we show the colour evolution of the Elliptical and Sbc galaxies in the \baosample as a function of redshift. We further show the \gold stellar locus for main-sequence stars. This highlights the possible colour confusion between galaxies and stars in our sample in the first redshift bins. Nonetheless, these regimes are where the morphological star/galaxy separation quantity in DES will work better. 

\subsection{Star contamination}
\label{sec:starcontamination}

Several studies have shown that star contamination can significantly affect the measurement of clustering of galaxies if not taken care of. On the one hand, they can damp the signal-to-noise ratio of the BAO signal~\citep{carnero2012}, but they can also introduce spurious power at large scales~\citep{ross11}, which could mimic the effect of primordial non-gaussianities. However, the effect on the angular position of the BAO is negligible. Therefore, some contamination level is allowed, especially since residual stellar contamination in the clustering is removed using the mitigation scheme presented in \secref{lsssys}.

In this work, as explained in \secref{baosample}, we use the \sofmash classifier to select secure galaxies, which should be a reliable star-galaxy separator in the range of $i=[19,22.5]$, with expected contamination in these magnitude ranges below $2\%$~\citep{y3gold}. Another effect is the possible obscuration of galaxies due to stars in the foreground. This effect is taken care of by the \gold foreground mask, which is applied to the \baosample and should account for the most obscured regions around stars; nonetheless, we can always expect that for low-brightness surface galaxies, this effect might be important, especially as we go to higher redshifts. 

We estimate the star contamination level by looking at the purity of the sample, defined as:
\begin{equation}
\label{eq:purity}
     \mathrm{Purity \ [\%]} = \left(1-\frac{\mathrm{n_{stars}}}{(\mathrm{n_{stars}}+\mathrm{n_{gals}})}\right) \times 100
\end{equation}
\noindent where $\frac{\mathrm{n_{stars}}}{(\mathrm{n_{stars}}+\mathrm{n_{gals}})}$ is the contamination.

\begin{figure*}
    \includegraphics[width=\linewidth]{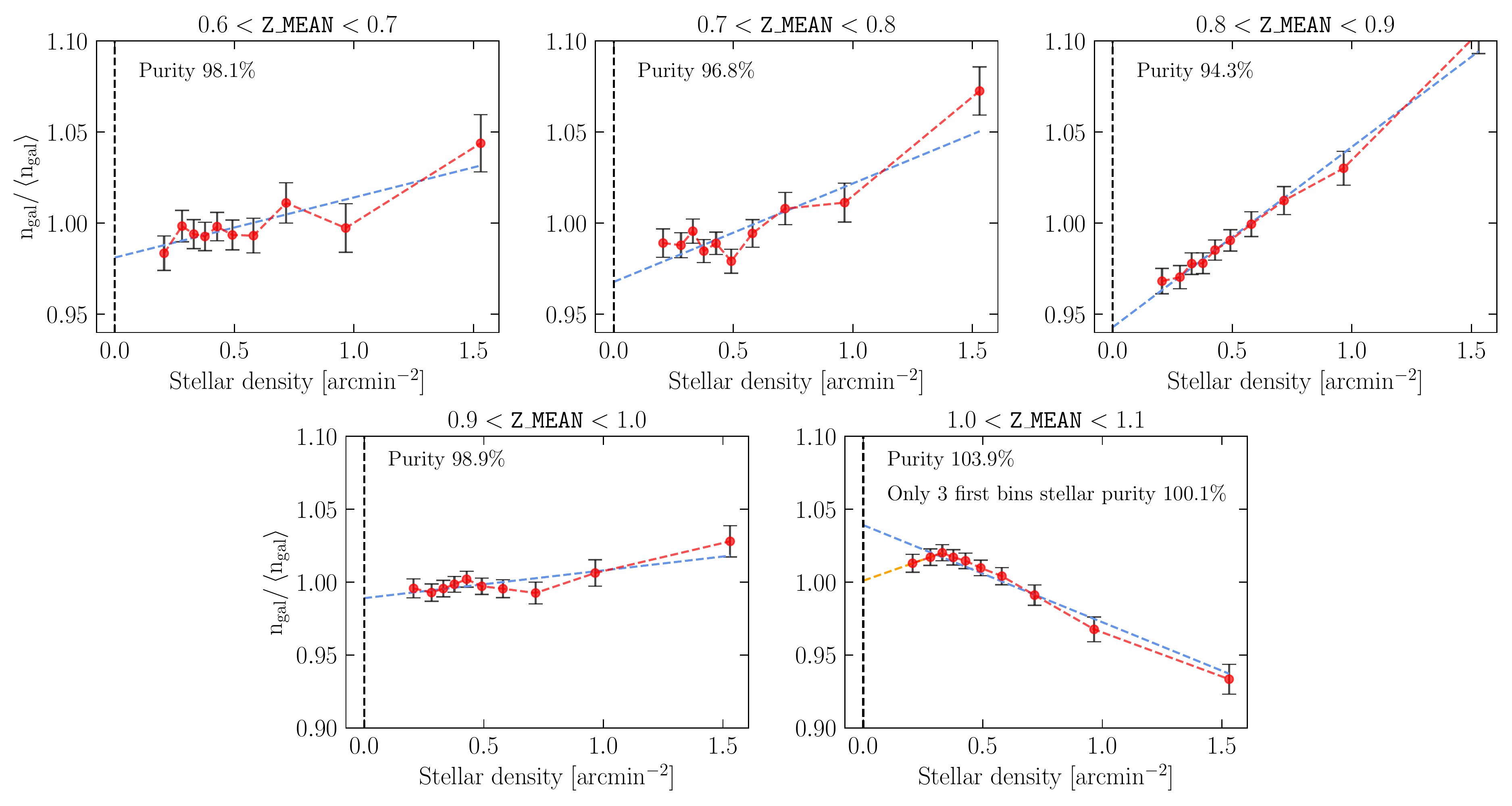}
\caption{Normalized density versus stellar density (in red). Apart from the last redshift bin, the number of objects in the \baosample increases with stellar density, expected if we consider contamination by stars. In the last bin, we see the effects of stellar obscuration, and therefore, the purity estimate is not reliable. Nonetheless, it is in this redshift bin where we expect less contamination. In blue, the fit to the distribution. Extrapolating to Stellar density = 0, we can infer the purity of our sample (Equation~\ref{eq:purity}). In the last redshift bin, we also show the linear fit, including only the first three points.}
    \label{fig:stars_1}
\end{figure*}

We start by applying the same algorithm as in Y1~\citep{baosample}. There we measured the galaxy density in the \baosample as a function of stellar density and extrapolate the relation to where stellar density = 0. This way, we can infer the sample purity by looking at the sample density in the absence of stars with respect to the mean density. We do this analysis separately for each tomographic bin to assess the sample's purity as a function of redshift. The result of this analysis can be found in \figref{stars_1}.

\begin{figure}
    \includegraphics[width=\linewidth]{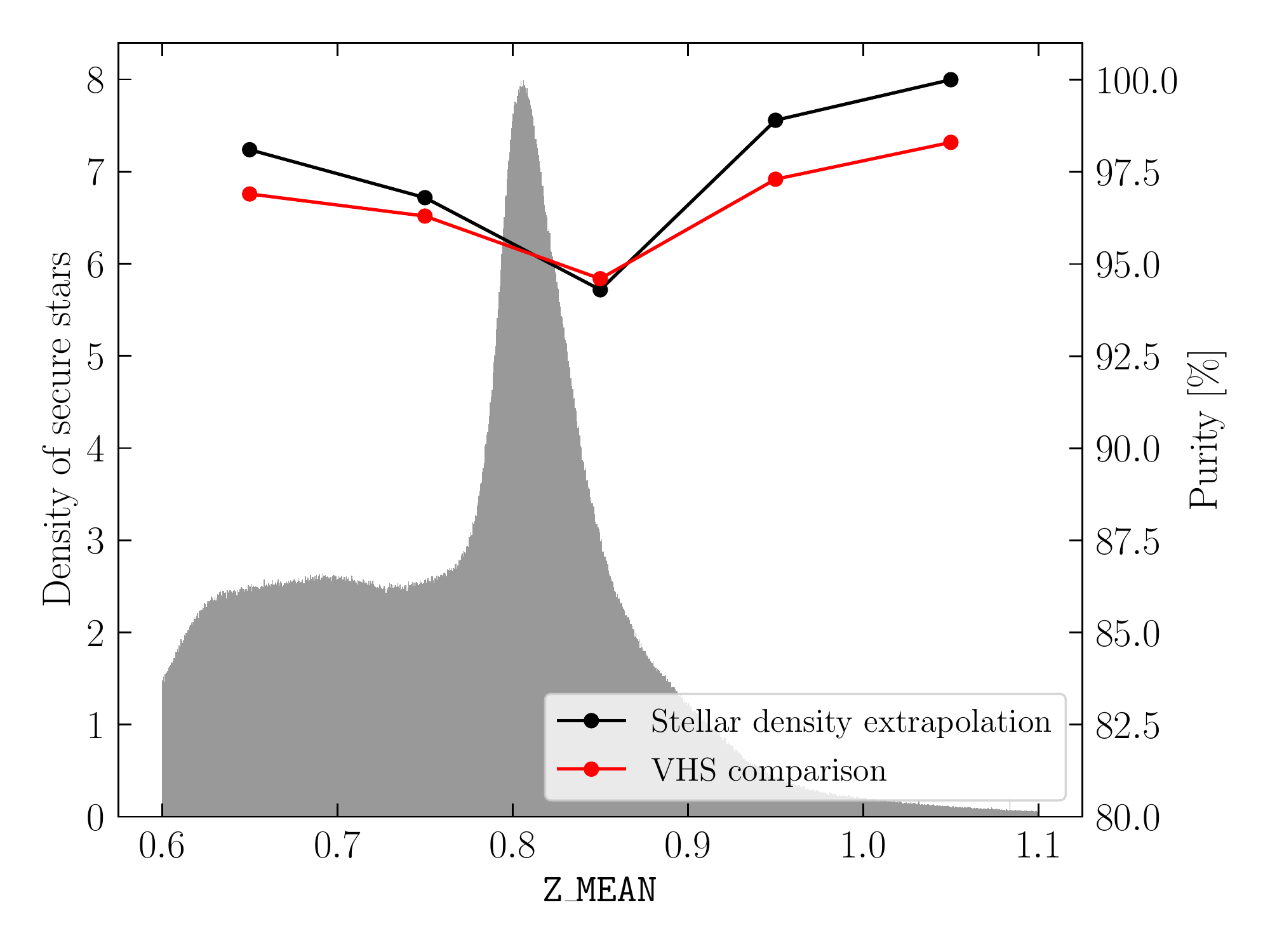}
\caption{In gray and on the left y-axis, the normalized density of secure stars if they were treated as galaxies that pass the \baosample colour cut versus \dnf \pz. The peak at \pz $\approx 0.8$ comes from all the stars with colour $r$-$i\gtrsim 1.3$. In lines and on the right y-axis, estimates of galaxy purity by two independent methods as a function of \pz. In black, we show the estimate based on the galaxy density as a function of stellar density, and in red, the purity based on the matched DES-VHS sample. Both estimates agree well and also with the trend seen in the gray distribution.}
    \label{fig:photoz_valid_magnitude_comparison}
\end{figure}

The contamination is small but somewhat larger than the average for \gold. The highest contamination is found in the second and third bin ($\mathrm{photo-}z \ =[0.7,0.9]$). This is confirmed in~\figref{photoz_valid_magnitude_comparison}. Here we show the \pz distribution if we assume that secure stars are galaxies, given the same colour cut applied to the \baosample. We can clearly see the distribution peaks in this redshift range. This is because of all stars will colours larger than $r$-$i$ $\gtrsim 1.3$, for which the best \pz will be that of the galaxies in the turn of the galaxy colour locus ($r$-$i$ $\approx 1.3$, $i$-$z$ $\approx 0.5$, cf. Figure~\ref{fig:colorcomp}). Consequently, for those stars the likelihood of the \pz fit worsen as they move away of the turn.

The contamination is probably overestimated due to obscuration effects in the borders of field stars. As already commented, this effect might be important in the highest redshift bins, and we see this effect in the last redshift bin ($\mathrm{photo-}z=[1.0,1.1]$). Here we expect very little contamination, but different from the other bins, we see a negative trend in galaxies' density as a function of stellar density. Internal studies ruled-out the possibility that an over-agressive star-galaxy separation was causing this effect. This warns us that the foreground mask developed for the \gold might be insufficient at the faint end and also, that purity might be over-estimated in the second to last redshift bin. Nonetheless, we expect this effect to have a negligible effect on the BAO measurement; likewise, we treat this effect on the mitigation of observing conditions in \secref{lsssys}.

To confirm our estimates, we further estimate the star contamination by comparing the \baosample with a different star/galaxy separation scheme, done in a subsample that combines DES photometry with near-infrared (NIR) data. It has been shown in~\citet{y1sgsep} that a combination of optical plus NIR data is excellent to discriminate between stars and galaxies. Therefore we match the \baosample to the DR4 Vista Hemisphere Survey~\citep[VHS]{vhs}. The overlap is $\roughly 2000$ square degrees, and Our sample contains 987254 of their sources. Unfortunately, VHS is only complete in DES up to $i \roughly 21$ so that we can expect some selection bias, especially at high redshifts. We estimate the percentage of stars in the \baosample following the colour separation seen in Figure 10 from~\citet{y3gold}. Therefore, we define as stars those sources with $J-Ks < 0.2 \times (g-i)+0.55$. Applying this cut to each redshift bin in the \baosample, we find the following purity levels: 96.96\%, 96.3\%, 94.6\%, 97.3\%, and 98.3\%. They are very similar to the ones found with the previous method, as can be seen in \figref{photoz_valid_magnitude_comparison}. In this case, the incompleteness of VHS at the fainter end is not important because it is here where less contamination is expected.

\section{Analysis blinding procedure}
\label{sec:blinding}

To avoid confirmation bias, DES follows strict blinding procedures. In general, we apply the same blinding guidelines for all the LSS samples. Here, the main rule is to avoid the calculation and visual inspection of the angular correlation function, $\omega(\theta)$, or the angular power spectrum before estimating the final cosmological constraints. Clustering statistics are only allowed to be calculated following these rules:

\begin{itemize}
    \item  Clustering measurements for any sample are allowed to be produced for a 10\% sub-sample of the area only, in up to 3 angular bins.
    
    \item Bias values can be obtained for these measurements, only using the halofit prediction of $\omega(\theta)$ for a fixed cosmology. These bias measurements are only meant to inform either the production of mocks or the forecast to optimized science analysis but not for parameter estimations.
\end{itemize}


In addition, we apply the blinding rules applied in Y1:

\begin{itemize}
    \item We cannot over-plot theory and data until the catalogues are frozen and all blinding tests have been approved. 
    
    \item No maximum likelihood values of any fits to data vectors will be reported until the catalogues are finalized. The width of a confidence interval may be, as well as the shape of the likelihood, as long as it is always centred on a fiducial value.
\end{itemize}

\subsection{Blind galaxy bias}
\label{sec:blindbias}

Based on these guidelines, we estimate the galaxy bias for the \baosample. These values are needed to construct the log-normal mocks used in the mitigation of observational systematic effects explained in~\secref{lognormalmocks}. We measure $\omega(\theta)$ for 3 angular bins from $\theta \roughly 0.58$ to $\theta \roughly 0.92$ degrees in a 10\% sub-sample, selecting a list of consecutive \healpix pixels, randomly selected within the footprint, but with preliminary weights defined from the whole footprint. We estimate a Gaussian+shot noise covariance matrix with input $C_\ell$ evaluated at MICE $\Lambda$CDM cosmology~\citep{mice,mice15}. First, we fit $\omega(\theta)$ with a covariance with galaxy bias $b=1$ in all redshift bins; then, we obtain the first set of temporary galaxy biases that are used to estimate the final covariance matrix. With this new covariance, we obtain the stated galaxy bias per redshift bin. In the process, we include the \nz's from \secref{photoz} and a first set of weights from the mitigation of systematics (using the Y1 \baosample bias as starting point). `Blind' galaxy bias values are given in \tabref{bao_properties}. 

\begin{figure*}
    \includegraphics[width=\linewidth]{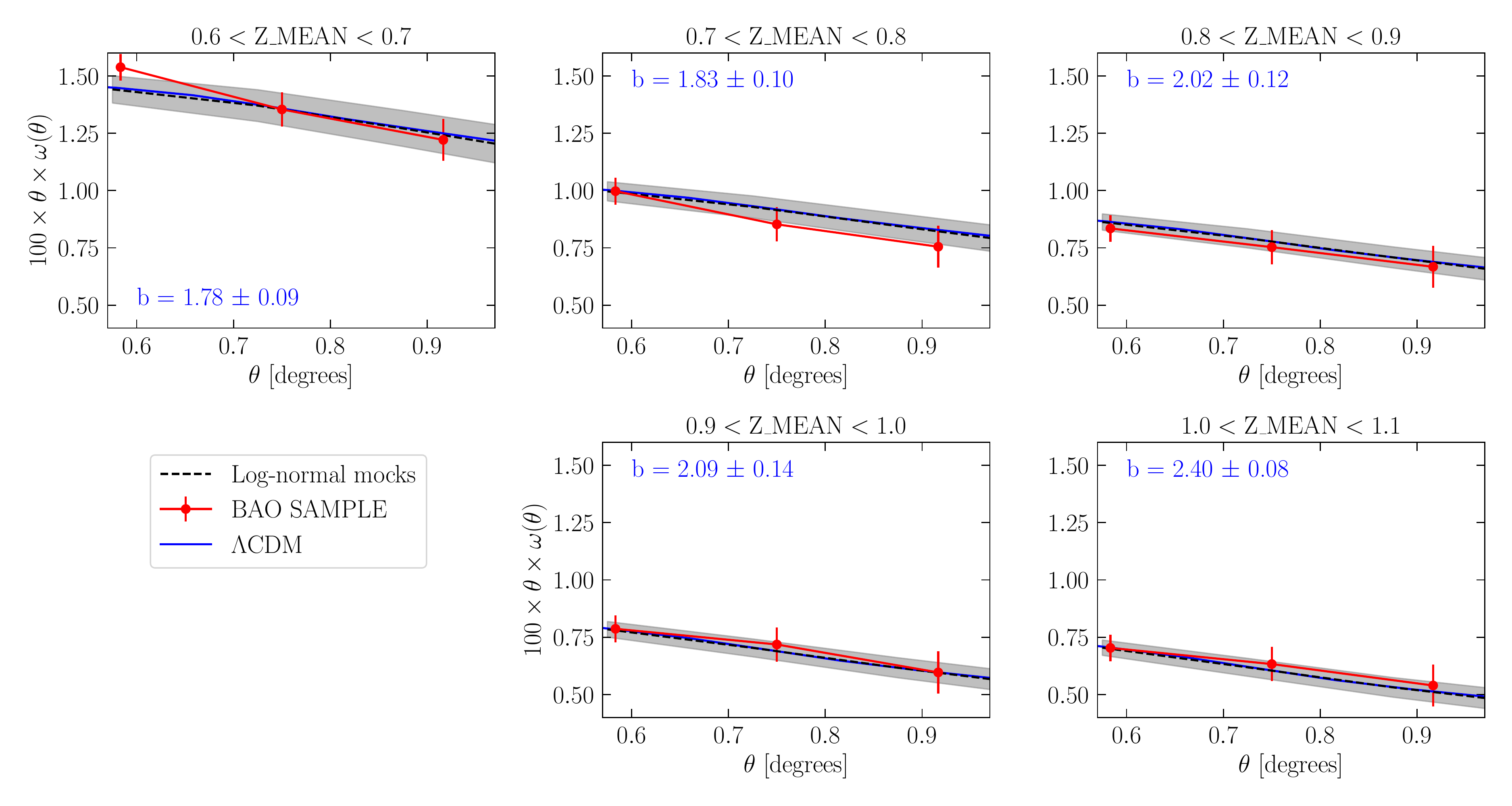}
\caption{In red is $100 \times \theta \times \omega (\theta)$ for the \baosample. These are the only data points allowed by the blinding procedure to be inspected. These 3 points are used to estimate a ``blind galaxy bias'' used to construct the log-normal mocks in~\secref{lognormalmocks}. In blue, the best fit prediction with fixed MICE $\Lambda$CDM cosmology, with $\Omega_{\mathrm{M}}=0.25$, $h=0.7$, $\Omega_{\mathrm{b}}=0.044$ and $n_{\mathrm{s}}=0.95$~\citep{mice}. The fit is done using the full covariance between points. In grey, the average $\omega (\theta)$ from the log-normal mocks.}
    \label{fig:blindbias_1}
\end{figure*}

In \figref{blindbias_1} we compare the \baosample $\omega (\theta)$ with the average clustering signal from the log-normal mocks (see~\secref{lognormalmocks}) and the theoretical model with the fitted galaxy biases.  

\section{Mitigation of Observational Systematic Effects}
\label{sec:lsssys}

This paper is based on observations taken during three years of operations at the Blanco telescope in Chile. On average, each position on the sky was observed four times (excluding the Supernova Fields), although the scatter is large, with some regions observed once, with other regions observed up to 10 times (cf. Figure 2 from \citealt{dr1}). Even though we impose for the \baosample that we have observed at least once in each band, the heterogeneous survey strategy implies fluctuations in seeing, exposure time, sky brightness, photometric calibrations, and other survey conditions, that, if not treated correctly, can imprint non-cosmological clustering in the density field. 

To correct these effects, we apply the iterative decontamination method presented in~\citet{elvinpoole}, already applied to the Y1 lensing samples. In Y3, we decided to apply the same methodology to all clustering catalogues, including the \baosample, \redmagic~\citep{redmagic} and \maglim~\citep{maglim} samples. We have called the method \textit{Iterative Systematic Decontamination} (ISD) method.

Details about the ISD method and results for the \redmagic~and \maglim\, samples are given in~\citet{monroy}. In this paper, we explain the methodology and document the results for the \baosample only.

\subsection{ISD Method}

The number density of galaxies is expected to fluctuate with the survey's imaging quality, both due to fluctuations in the noise and due to limitations of the selection process. The method developed in ~\citet{elvinpoole} consists of assessing how much the galaxy density varies with respect to a given survey property, given that the natural variations in number density do not correlate with survey properties. When a significant relationship is found, for example, when the galaxy density increases or decreases with seeing or airmass, a weight is assigned to each galaxy as a function of the observing condition's value at its position to correct for this fluctuation. In a real case, we have hundreds of survey conditions correlated with each other and for several bands. For this reason, a robust and automatic methodology is needed to correct for observing condition fluctuations. 

The ISD method is an iterative method. It starts by estimating the significance of the galaxy density versus each survey property. The method uses \healpix maps for each survey property as well as to estimate galaxy densities, where we consider the effective area of the pixels given by the \gold \var{FOOTPRINT} map~\citep{y3gold}. 

To measure the significance, we start by minimizing $\chi^2_{\mathrm{model}}$ for each \textit{ith} survey property map (SP map), where the model is:

\begin{equation}
	 \frac{n_{i}}{\langle n  \rangle} = m \cdot s_i + c \ ,
\label{eq:linearfit}
\end{equation}

with $n_{i}$ the number of galaxies as a function of the SP map and $\langle n  \rangle$ the average. Finally we estimate a $\Delta \chi^2$ as:

\begin{equation}
\Delta \chi^2 = \chi^2_{null} - \chi^2_{model} \ , 
\end{equation}

\noindent between the best-fit linear parameters and a null test with $\frac{n_{i}}{\langle n  \rangle} = 1$. The covariance used to calculate these $\chi^2$ values is given by the dispersion of the same galaxy density - SP map relation measured on the uncontaminated log-normal mocks (see~\secref{lognormalmocks}).


If $\Delta \chi^2$ is actually significant or not, will depend on the noise properties of the sample. Therefore, in order to assess the degree of significance of each SP map, we compare the $\Delta \chi^2$ obtained from the data with the probability distribution of $\Delta \chi^2$ obtained from a set of galaxy mocks, representing the same numbers, areas, and redshift distributions as the data. From the mocks probability distribution we define $\Delta \chi^2 (68)$ as the impact degree below which $68 \%$ of the $\Delta \chi^2$ from the mocks are. Finally, we define the significance of each map SP map as:
\begin{equation}
	    S_{1D} = \frac{\Delta \chi^2}{\Delta \chi^2 (68)} \ .
\end{equation}

\noindent The second step is to remove the galaxy density vs survey property relation for a given map by weighting galaxies according to the inverse of the linear relation fit for that map. At each iteration we use the SP map with the highest significance, $S_{1D}$, for this step. Finally, we move to the next iteration, once galaxies have been weighted (corrected). 

The method ends when all the $S_{1D}$ are below a given, user pre-defined threshold, $T_{1D}$. The choice of $T_{1D}$ should balance the levels of residual contamination and of over-correction. The final weight is defined as the multiplication of all the weights computed in each iteration and is normalized such that $\langle w \rangle = 1$ (average over footprint).

In general, it is not necessarily true that the map to weight for at a given iteration \textit{j} will be the map with \textit{j}-th highest $S_{1D}$ at iteration 0, because the existing correlations between SP maps make weighting for a given map to have an effect on others' significance. This correlation is dealt naturally by the ISD method. 

\begin{figure*}
    \includegraphics[width=0.8\linewidth]{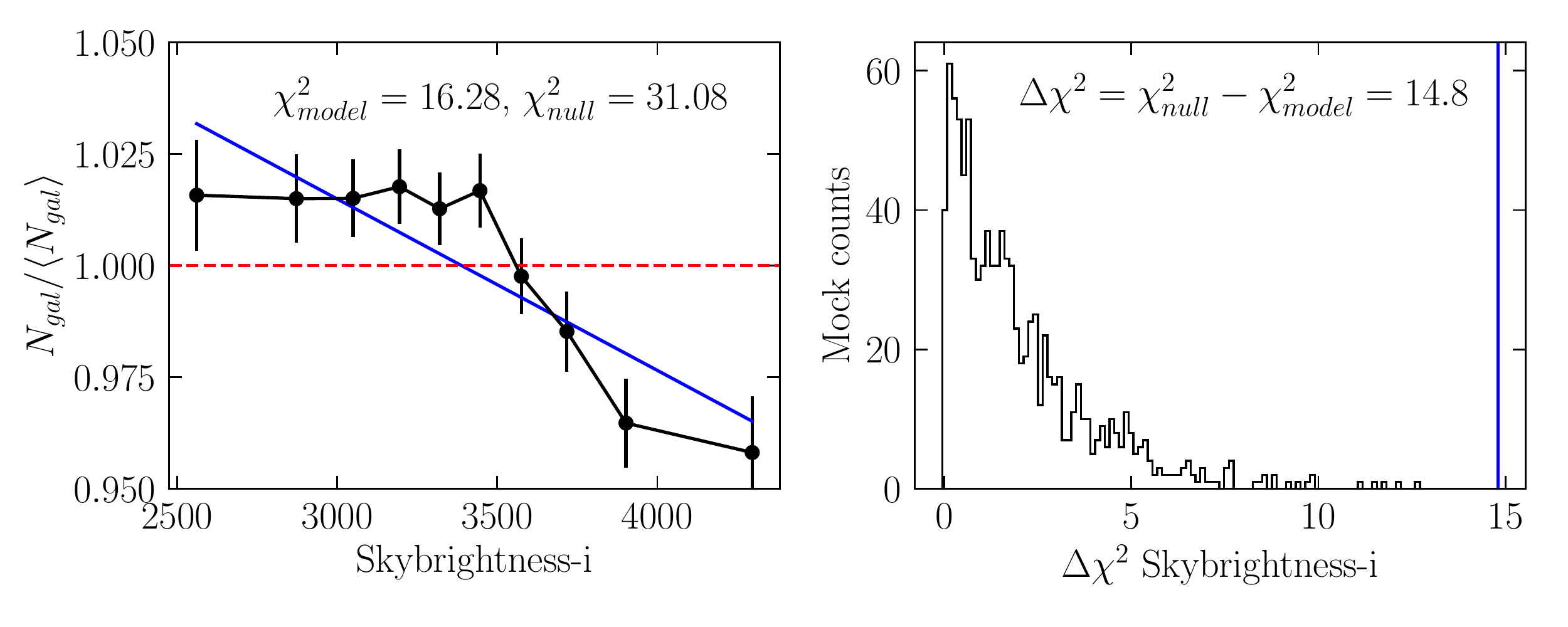}
\caption{Summary of one iteration in the ISD method. First, we calculate the minimum $\chi^{2}$ value given a linear polynomial fit, and also, by a constant relation. We estimate $\Delta \chi^2$ between both and compare it with the probability distribution of $\Delta \chi^2$ in a set of mocks. If  $\frac{\Delta \chi^2}{\Delta \chi^2 (68)}$ is above a given threshold ($T_{1D}$) and it is the largest from the list of SP maps available, we will use it to create weights following the linear relation found. We repeat this procedure for the remaining maps until all are below $T_{1D}$. In the case of the \baosample, we chose a $T_{1D}=4$, equivalent to a confidence level $\geq 99\%$.}
    \label{fig:methodsys}
\end{figure*}

Each iteration step of the ISD method is summarized in \figref{methodsys}, given the results for the first redshift bin in the first iteration. On the left, we see the galaxy density as a function of sky-brightness in the \iband; as we go to higher sky-brightness regions, the density of galaxies decreases. To assess if this is significant or not, we compare $\Delta \chi^2$ from the data to that from 1000 mocks (right-hand side plot). In this case, this relation is the most significant of all the SP maps, and therefore, we will correct the galaxy density by this survey property first and continue to the second iteration.

Recently, other methods have been proposed for the correction of observational systematics in DES Y3 data~\citep{ENET,otweight,wagoner2020}. Moreover, ISD has also been used with alternative configurations, such as using a PCA of the SP maps. A comparison of some of these methods and configurations applied to the \var{REDMAGIC} sample is shown in~\citet{monroy}. For the \baosample, none of these methods were available by the time of the freeze of the sample that followed the blinding procedure. The differences between these methods, though small, mostly affect the clustering amplitude but they have a marginal impact on the position of the BAO peak. We checked this impact by comparing the results from our fiducial weights with those from PCA~\citep{monroy}, observing negligible differences.

\subsection{Input ingredients}

Before running the ISD method, we need to define several inputs. For example, we need to define the threshold $T_{1D}$ below which we demand all SP maps to be below at end of the run. Furthermore, we have assumed a linear fit to model the galaxy density vs. SP map relationship, but nothing prevent us from using higher-order functions. Nonetheless, internal tests have shown that a linear fit is more than sufficient to cope with galaxy density variations for all survey conditions, at least for the level of precision we need for galaxy clustering in the \baosample. Moreover, higher order fits could lead to over-fitting.

\subsubsection{Observing condition maps}
\label{sec:spmaps}
DES produces hundreds of observing condition maps for each band in \healpix format. The definition of these maps are detailed in~\citet{y3gold}. For convenience, since several of these maps are correlated, we reduce the list to a set of 32 maps (8 in each $griz$ band). Details about the SP maps list reduction is given in \appref{rep_spmaps}. Besides, we also create a stellar density map (based on DES morphological classification) and include a galactic extinction map, in this case, the SFD98 map of~\citep{sfd98}.  Details about the creation of these maps are given in~\citet{y3gold}. Finally, the list of SP maps used for the \baosample is (weighted quantities use inverse-variance weights of the single-epoch photometric errors):

\begin{itemize}
	\item \texttt{AIRMASS} ($griz$): weighted mean airmass from all exposures.
	\item \texttt{SKYBRITE} ($griz$): weigthed mean sky brightness from all exposures.
	\item \texttt{SKYVAR\_UNCERT} ($griz$): weighted Sky variance with flux scaled by zero point.
	\item \texttt{FWHM\_FLUXRAD} ($griz$): Twice the average half-light radius from the sources used for determining the PSF.
	\item \texttt{FGCM\_GRY} calibration ($griz$): residual `grey' correction to the zero-point.
	\item \texttt{SOF\_DEPTH} ($griz$): galaxy depth at $10\sigma$ for \SOF, corrected for zero-points and galactic extinction.
	\item \texttt{SIGMA\_MAG\_ZERO} ($griz$): Quadrature sum of zero-point uncertainties.
	\item \texttt{T\_EFF\_EXPTIME} ($griz$): exposure time weighted by the effective time of observation.
    \item \texttt{STELLAR\_DENS}: density of stars.
    \item \texttt{SFD98}: extinction map from~\citet{sfd98}.
\end{itemize}

\subsubsection{Galaxy mocks}
\label{sec:lognormalmocks}

To assess if the density relations in the ISD method are significant or not, we rely on a set of mocks. Given the method's flexibility, we rely on the log-normal formalism of~\citet{coles91} to create a set of 1000 mocks for our analysis. We checked that 1000 mocks were sufficient. 

The method is the following: the first step is to log-normalize the real-space $C_\ell$'s (see equation 21 in~\citealt{xavier16}). After this transformation, $C_\ell$'s can be used to create gaussian fields for the matter overdensities. The generated log-normal overdensity maps are then masked and normalized to the input number of galaxies, so the total density is equal to the sample's density. Finally, we draw random galaxies following a Poissonian distribution to mock the shot-noise on our galaxy sample in the survey masks. The final products of these mocks are \healpix density maps. We can produce maps in any given resolution. Tests showed that an \nside=512 was sufficient without degrading the results.

In our analysis, we run the method twice: first to get a preliminary set of weights that we use to estimate the blind galaxy bias explained in \secref{blindbias}. Using these results, we are able to compute theoretical $C_\ell$'s matching the clustering amplitude, that are used both to feed the log-normal mocks and also the COLA mocks in~\citet{colamocks}. Finally, using the new theoretical $C_\ell$'s, we compute 1000 log-normal mocks to feed the mitigation of observational systematics.

Several internal tests showed that changing the fiducial cosmology, \nz, or galaxy biases did not affect the mitigation of systematic effects. This is expected since the general form of the clustering is not important here. 

Later in \secref{weightsvalidation} we will validate the weights over a set of ``contaminated'' mocks, contaminated with the same observational effects seen in the data. To do so, we will vary the number density according to the \baosample weights from the same Poisson noise distribution as the ``non-contaminated'' mocks, therefore, drawing non-contaminated and contaminated mocks from the same random distribution. This is slightly different from the method applied in~\citet{monroy} to the other LSS samples where the ``contaminated'' mocks are drawn from independent Poisson noise realizations.


\subsubsection{Choice of threshold}

Unlike the other LSS samples, for the \baosample, we use a $T_{1D}=4$, equivalent to a confidence level $\geq 99\%$. Our forecasts showed the choice of a strict or loose $T_{1D}$ has almost a negligible effect on BAO measurement (cf. \figref{alpha_diff}). Therefore, to avoid over-corrections and uncertainties propagation's in cosmological estimates, we selected a $T_{1D}=4$ as our choice. More details are given in \secref{weightsvalidation} and \appref{choicethreshold}. Likewise, the mitigation corrections are, at first approximation, flat. Therefore, any remaining systematic not corrected for will be absorbed by the galaxy bias. In the case of BAO measurement, it means that it will have a negligible effect in cosmology, independently of how much we correct. However, some caution must be taken in the case of studying primordial non-gaussianities or other large-scale observables. 

\subsection{Results for systematic mitigation}
\label{sec:mitigationresults}

\begin{figure*}
    \includegraphics[width=0.90\linewidth]{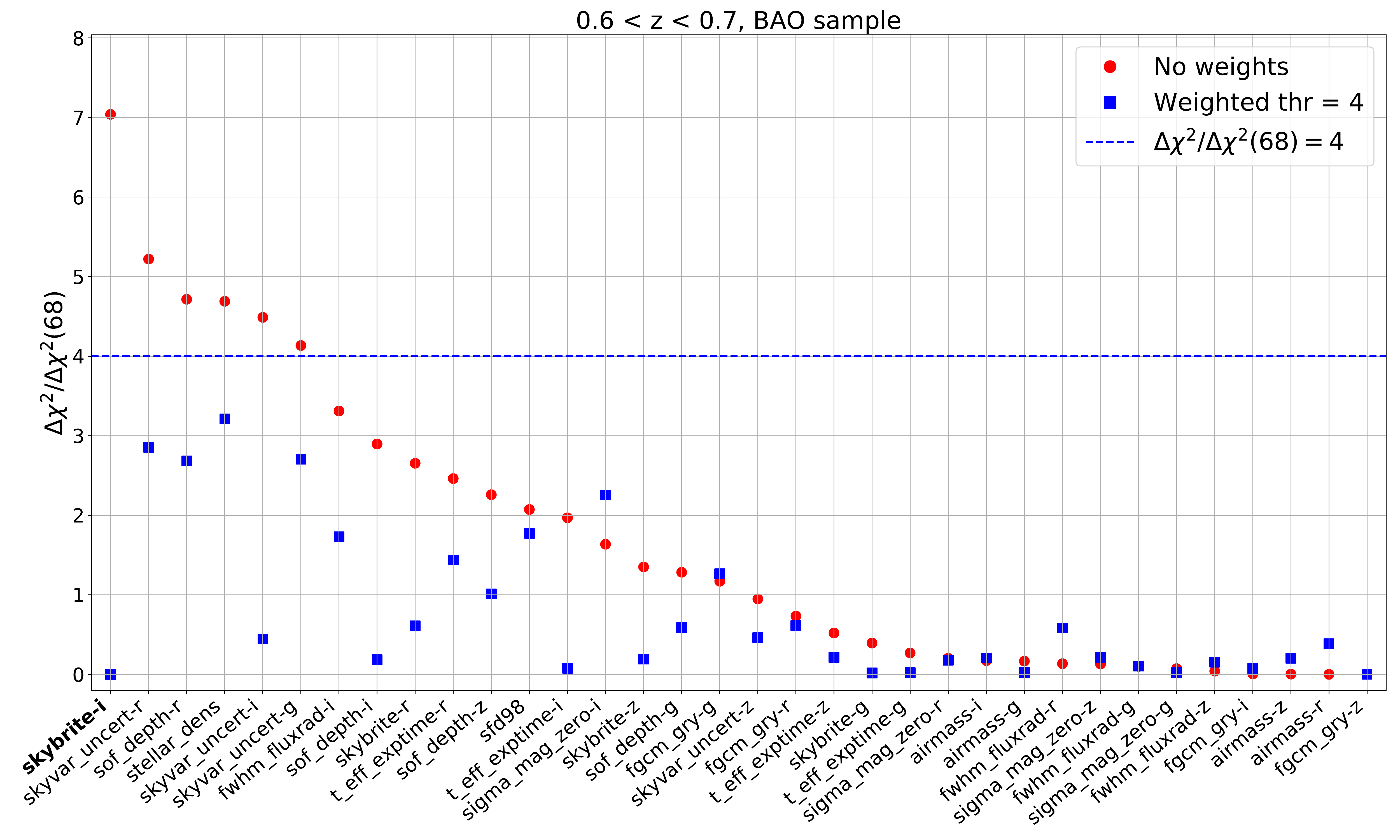}
    \includegraphics[width=0.90\linewidth]{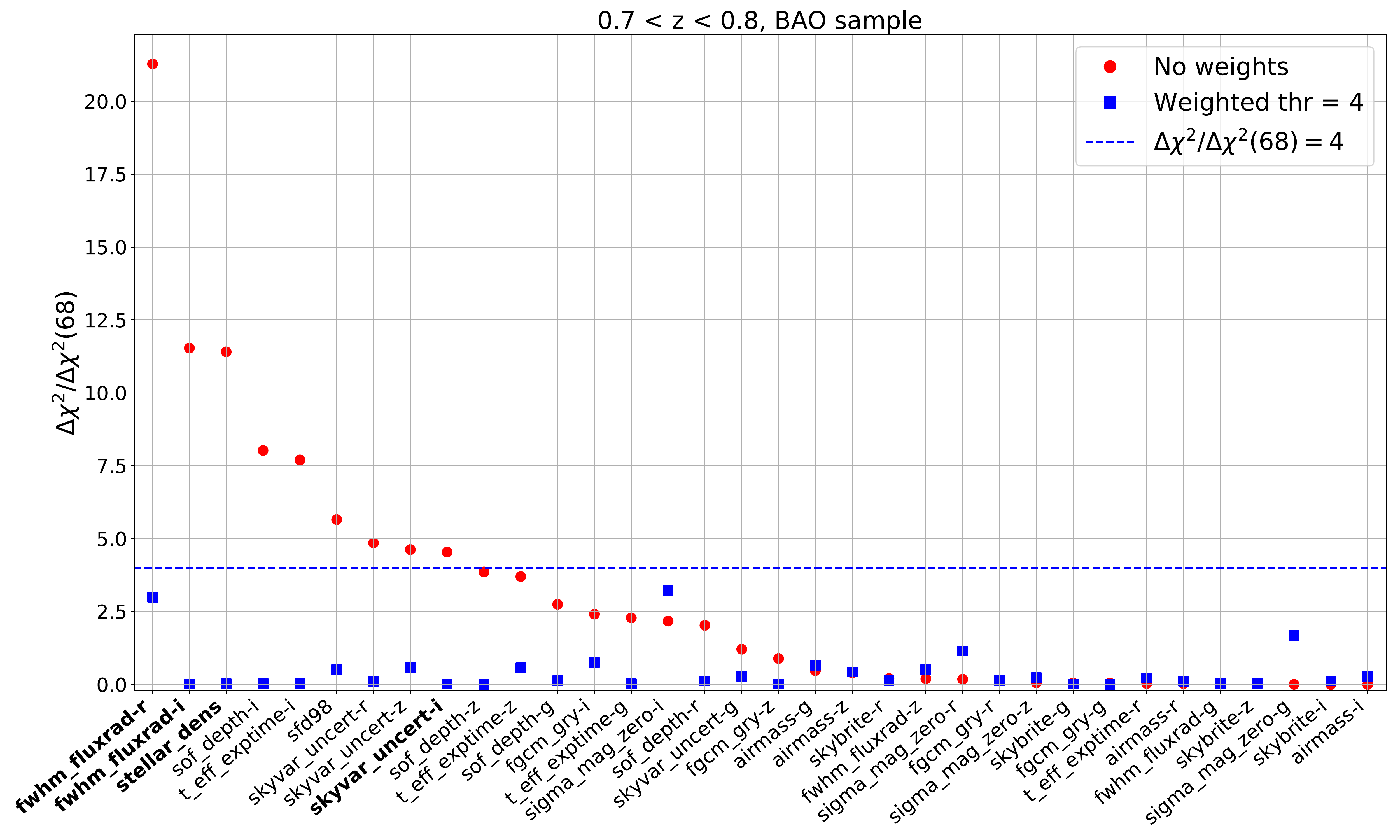}
    \caption{Significance of each SP map with respect to the \baosample for the first two redshift bins. Figures show the significance of each SP map before starting the method (in red circles) and also, after the method ends once all SP maps are under $T_{1D}=4$ (blue squares). The horizontal dashed line correspond to the significance threshold. The {\bfseries bold} labels on the x-axis list the SP maps that have effectively being used to create the final weights.}
    \label{fig:sigma_evol_1}
\end{figure*}

\begin{figure*}
    \includegraphics[width=0.90\linewidth]{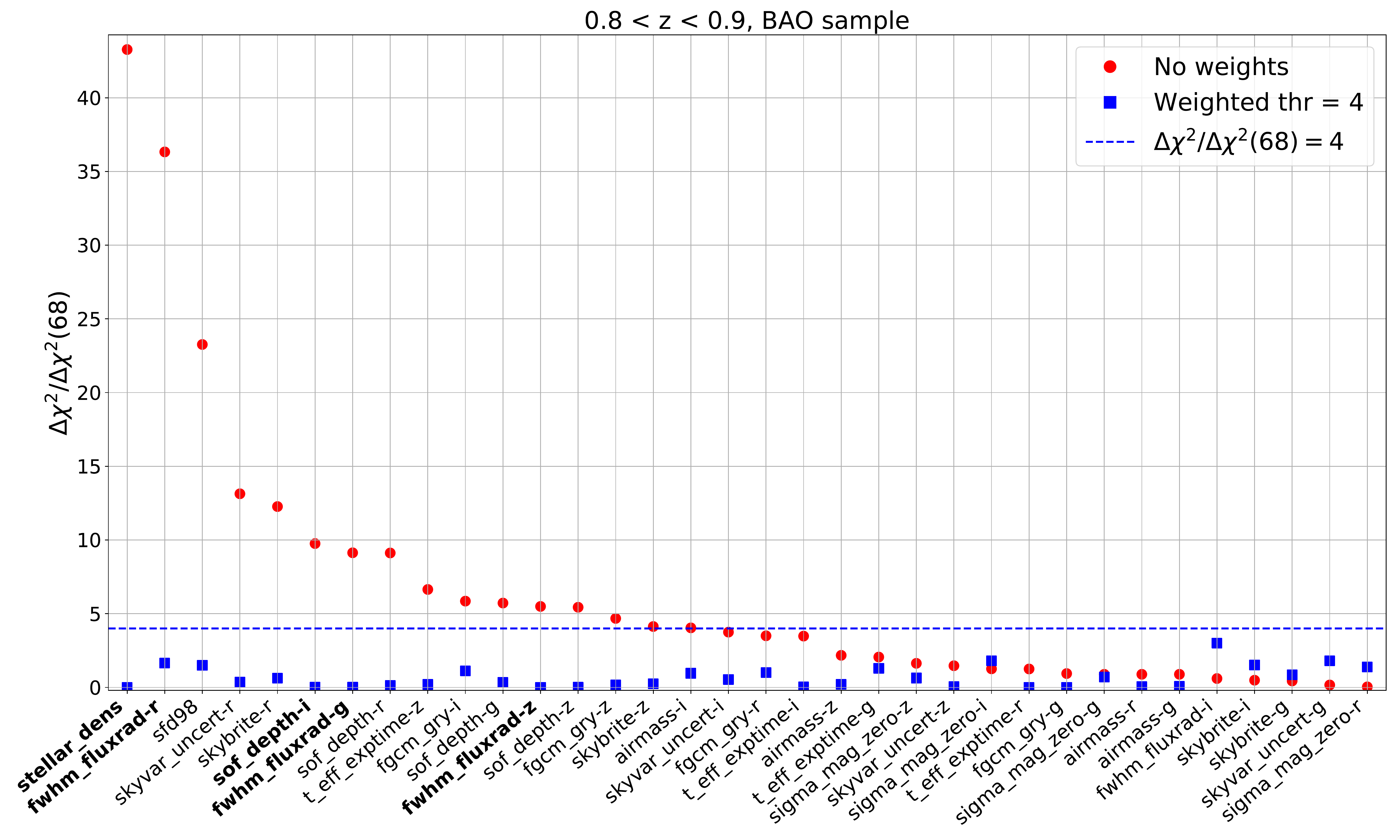}
    \includegraphics[width=0.90\linewidth]{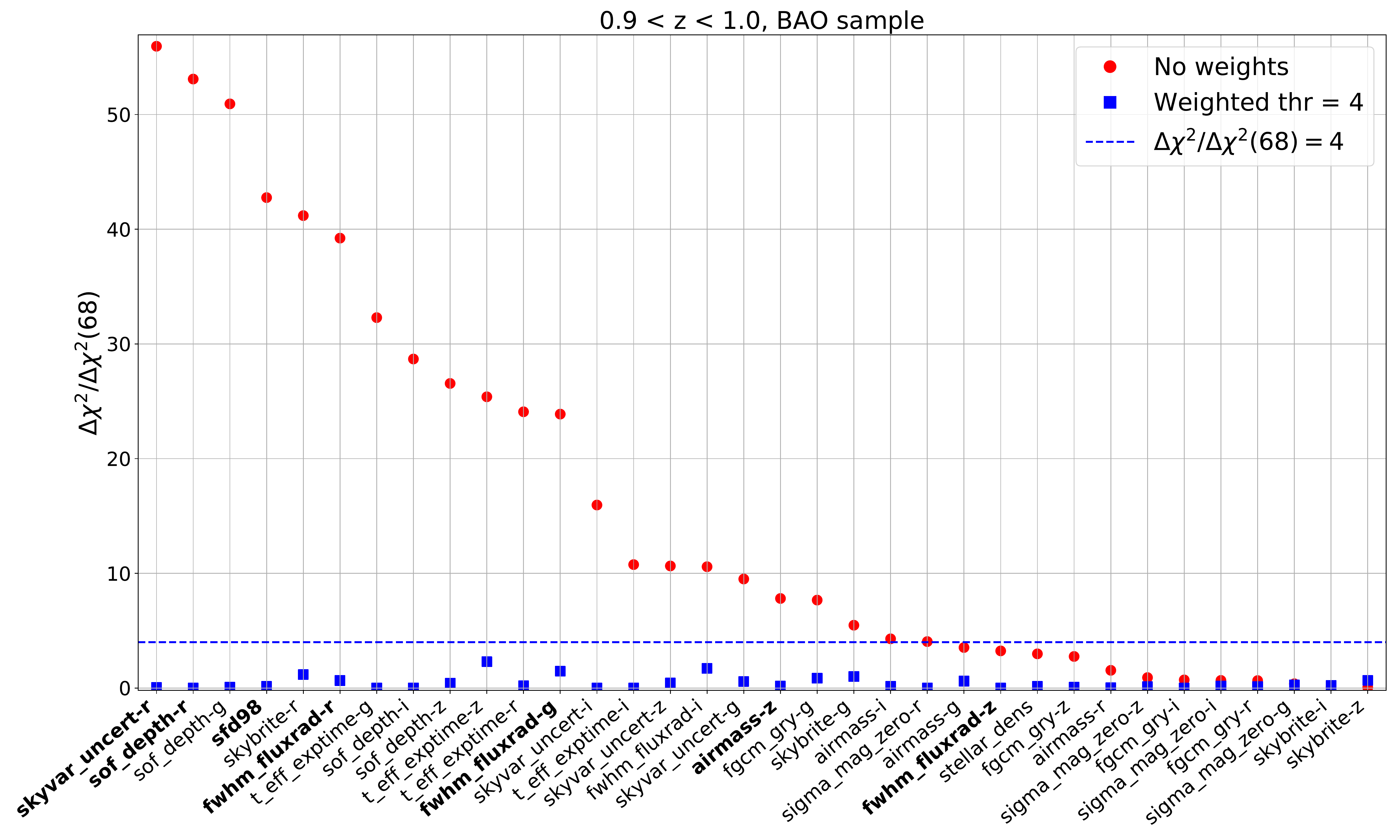}
    \caption{Significance of each SP map with respect to the \baosample for the third and forth redshift bin. Figures show the significance of each SP map before starting the method (in red circles) and also, after the method ends once all SP maps are under $T_{1D}=4$ (blue squares). The horizontal dashed line correspond to the significance threshold. The {\bfseries bold} labels on the x-axis list the SP maps that have effectively being used to create the final weights.  }
    \label{fig:sigma_evol_2}
\end{figure*}

We run the ISD method on the \baosample, with $T_{1D}=4$. In Table~\ref{tab:spmaps_baosample} we present the maps that we need to correct for in each redshift bin. We find that in most redshift bins, we need to correct by seeing and sky-variance uncertainty, and in some cases, by stellar density. In \figref{sigma_evol_1}, \figrefs{sigma_evol_2}{sigma_evol_3} we show the ordered list of maps before starting the iterative process, in decreasing order of significance. Also, in these figures, we show the significance level for all SP maps once we stop the process.

\begin{table*}
    \centering
        \begin{tabular}{ | p{2cm} | p{12cm} | }

        \multicolumn{2}{ |c| }{\baosample} \\ 
        \hline \pz bin & SP maps used to estimate the systematic weights given a tolerance in the method of $T_{1D}=4$ \\ \hline
         $0.6 < z < 0.7$ & skybrite-i  \\ \hline 
         $0.7 < z < 0.8$ & fwhm\_fluxrad-r, fwhm\_fluxrad-i, stellar\_dens, skyvar\_uncert-i \\ 
         \hline 
         $0.8 < z < 0.9$ & stellar\_dens, fwhm\_fluxrad-r, fwhm\_fluxrad-g, \\ & fwhm\_fluxrad-z, sof\_depth-i \\ \hline 
         $0.9 < z < 1.0$ & skyvar\_uncert-r, sfd98, fwhm\_fluxrad-g, fwhm\_fluxrad-r, airmass-z, \\ &  sof\_depth-r, fwhm\_fluxrad-z \\ \hline 
         $1.0 < z < 1.1$ & skyvar\_uncert-r, fwhm\_fluxrad-i, fwhm\_fluxrad-g, stellar\_dens, \\ & fwhm\_fluxrad-r, fwhm\_fluxrad-z, skyvar\_uncert-g, sfd98, airmass\_z, t\_eff\_exptime-z \\ \hline 
          
    \end{tabular}
    \caption{List of SP maps found to have impact on the \baosample sample at each redshift bin. The rows should be read from left to right in order of importance.}
    \label{tab:spmaps_baosample}
\end{table*}

\begin{figure*}
    \includegraphics[width=0.90\linewidth]{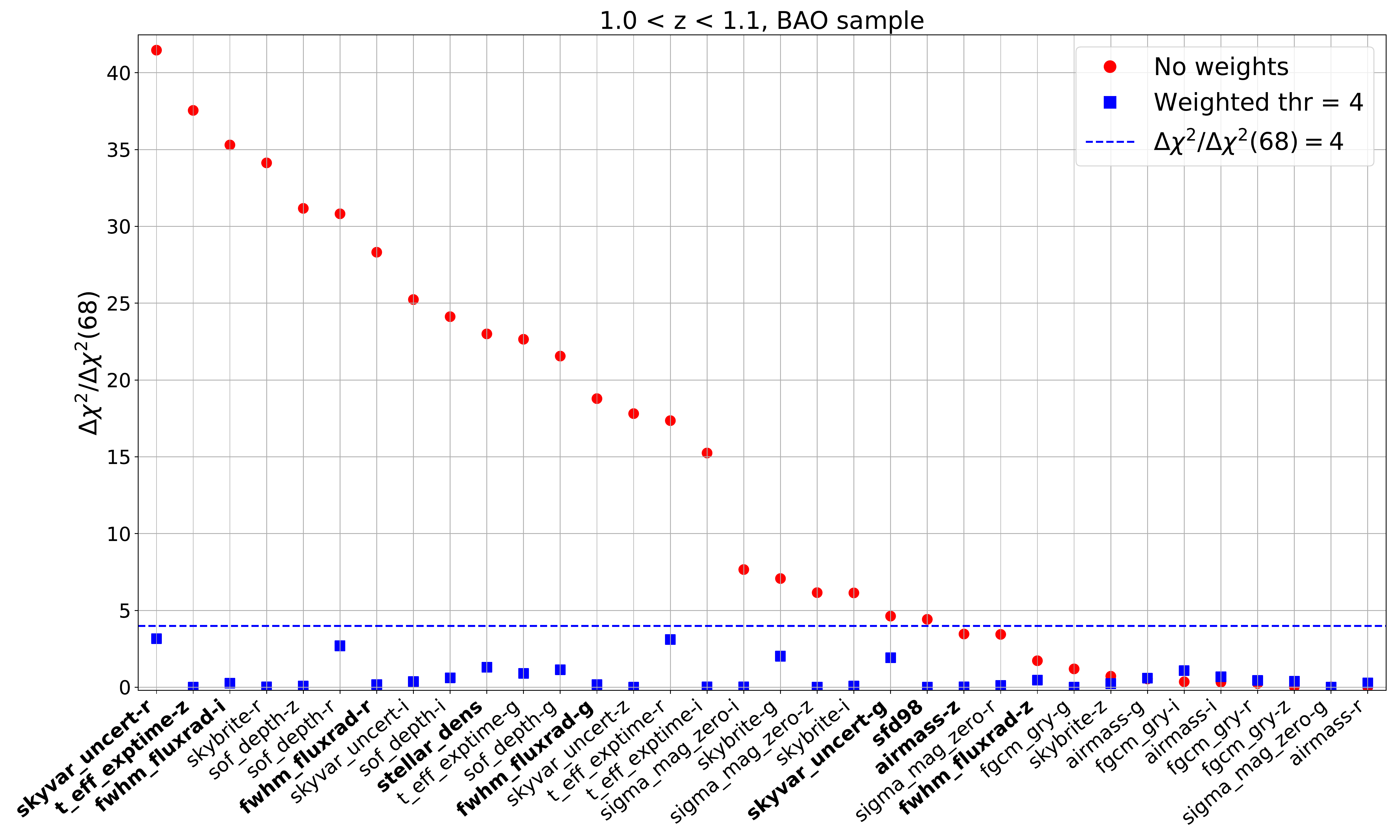}
    \caption{Significance of each SP map with respect to the \baosample for the last redshift bin. Figures show the significance of each SP map before starting the method (in red circles) and also, after the method ends once all SP maps are under $T_{1D}=4$ (blue squares). The horizontal dashed line correspond to the significance threshold. The {\bfseries bold} labels on the x-axis list the SP maps that have effectively being used to create the final weights. }
    \label{fig:sigma_evol_3}
\end{figure*}

\begin{figure*}
    \includegraphics[width=\linewidth]{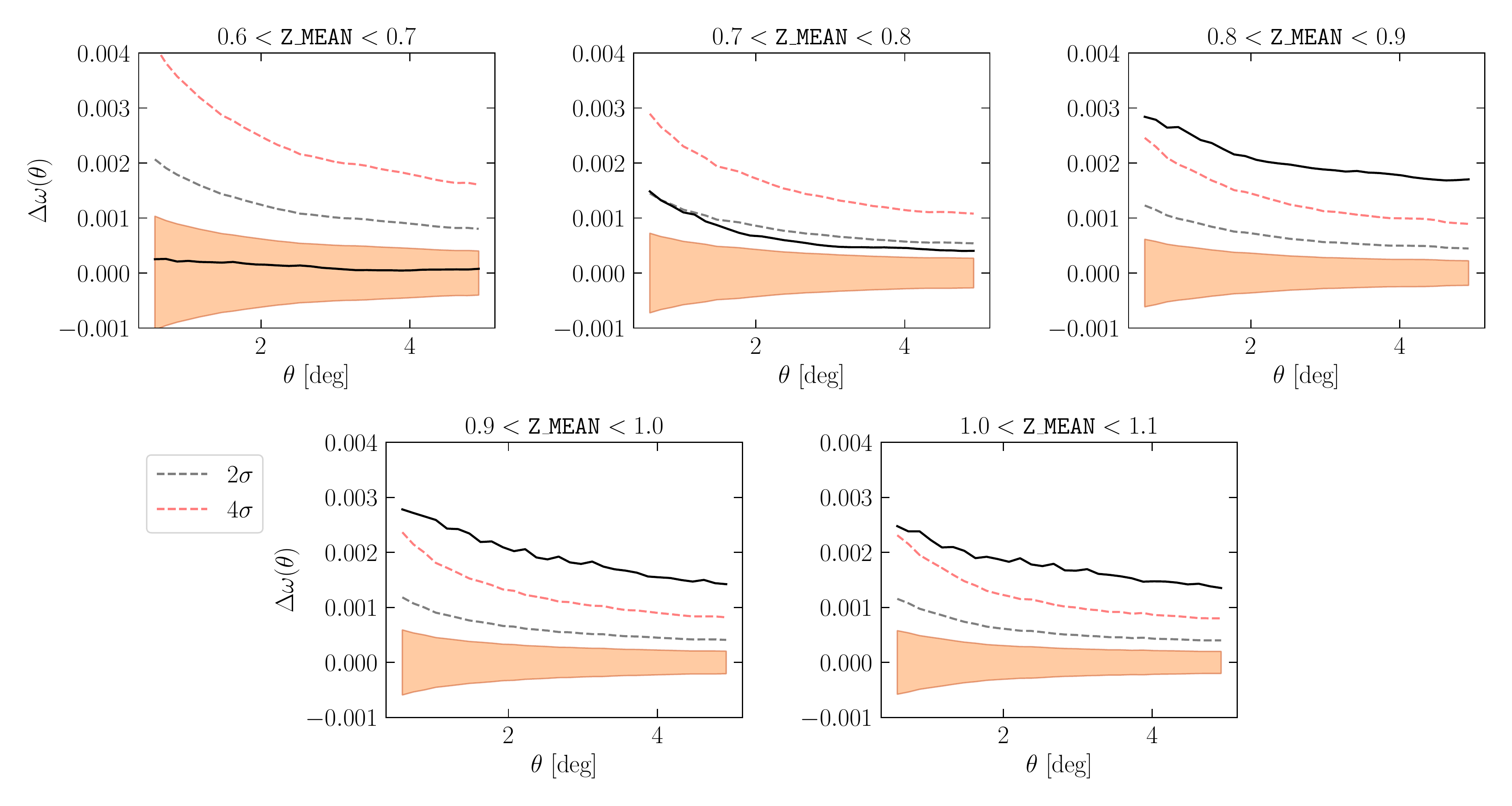}
\caption{Difference between $\omega(\theta)$ before and after applying the weights, in black. In shaded region, the statistical error calculated as the standard deviation of the 1000 log-normal mocks. In dashed lines the $2\sigma$ and $4\sigma$ error level.}
    \label{fig:deltatest_1}
\end{figure*}

Once we calculate the weights (see~\figref{deltatest_1}), we evaluate the impact on $\omega(\theta)$ by measuring the difference before and after applying weights, showed in~\figref{deltatest_1}. This comparison follows the blinding procedure described in \secref{blinding}. As we move to higher redshifts the correction increases, reaching a level which is several times the statistical error. At the start, this was a puzzling result, and an intense analysis was devoted to optimize the galaxy mask in order to reduce these levels. Still, after eliminating regions with the highest levels of variations in observing conditions, we did not find a significant reduction to propose a decrease in the survey area. In these tests, we concluded it was naturally occurring due to the broad area of the survey and to the larger list of observing conditions with respect to Y1. 

At this point is worth noting what happens if we do not correct at all by these effects in cosmological forecasts. In~\appref{choicethreshold} we apply the template-based method used by DES~\citep{y3mainbao} to recover the BAO scale, for both contaminated and uncontaminated log-normal mocks. We find that the template-based method is insensitive to these effects and we recover the true BAO scale in both cases. Therefore, the method is robust against the amplitude of the weights. In any case, since the 2-pt statistics (angular correlation function and power spectrum) can be used beyond the estimate of the BAO scale, it is worth applying the weights to measure the unbiased estimates.

Finally, after passing all the validation tests presented in the next section, we conclude the effect is well-characterized and, therefore, not prone to systematic errors beyond the $\omega(\theta)$ statistical error. 

\subsection{ISD method validation}
\label{sec:weightsvalidation}

We have performed several validation tests on the ISD method, designed to estimate the bias we are introducing in $\omega(\theta)$ after the de-contamination process. More details about the methodology are described in \citet{monroy}. There we validate the methodology for the other LSS samples, using a different list of SP maps (based on principal components). Here we present the results for the \baosample, using the list of 34 SP maps presented in \secref{spmaps}. It is worth noting that we tested the use of the principal components SP maps after unblinding but we did not find any major difference. 

\begin{figure}
 \includegraphics[width=\linewidth]{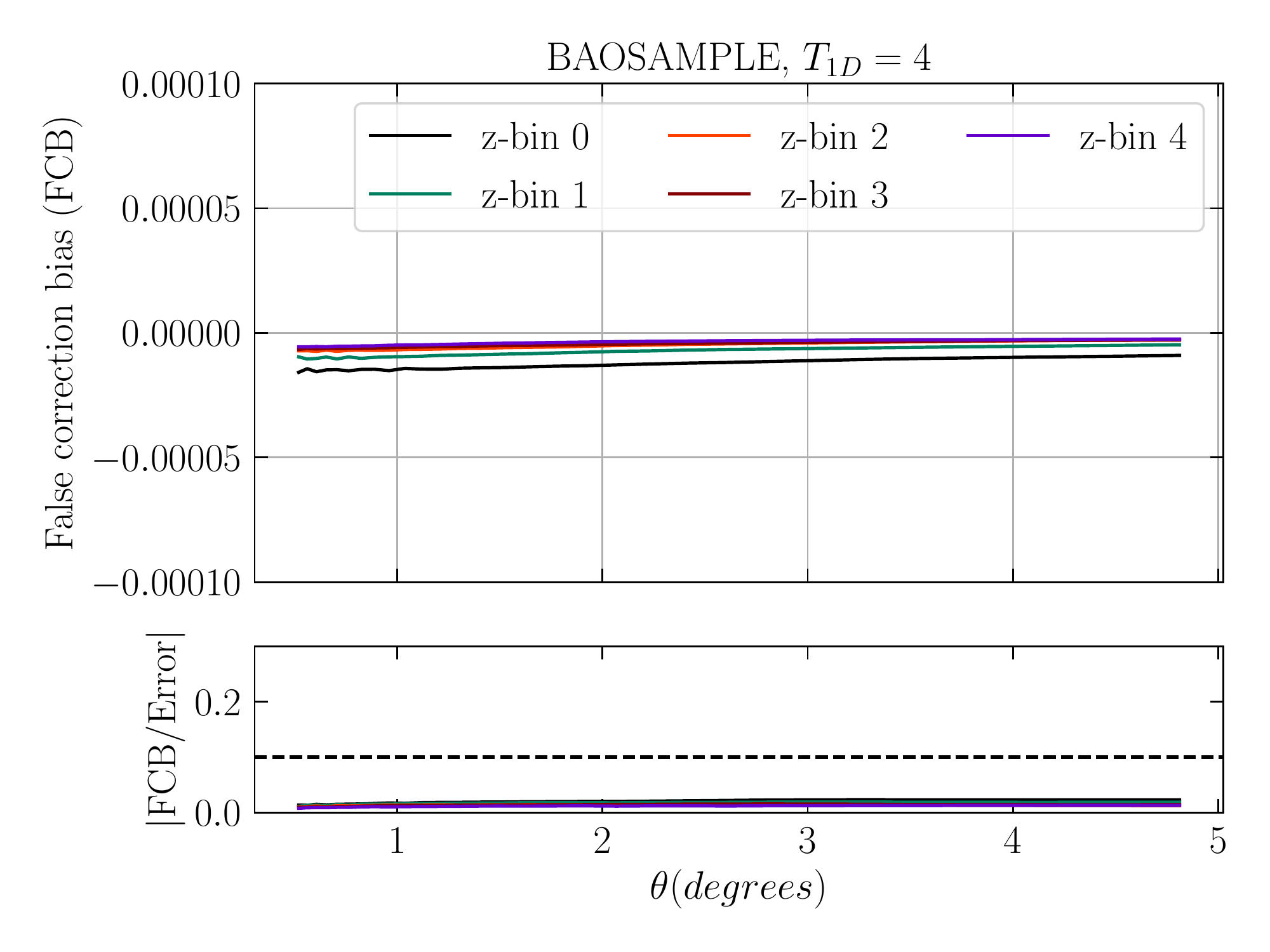}
 
    \includegraphics[width=\linewidth]{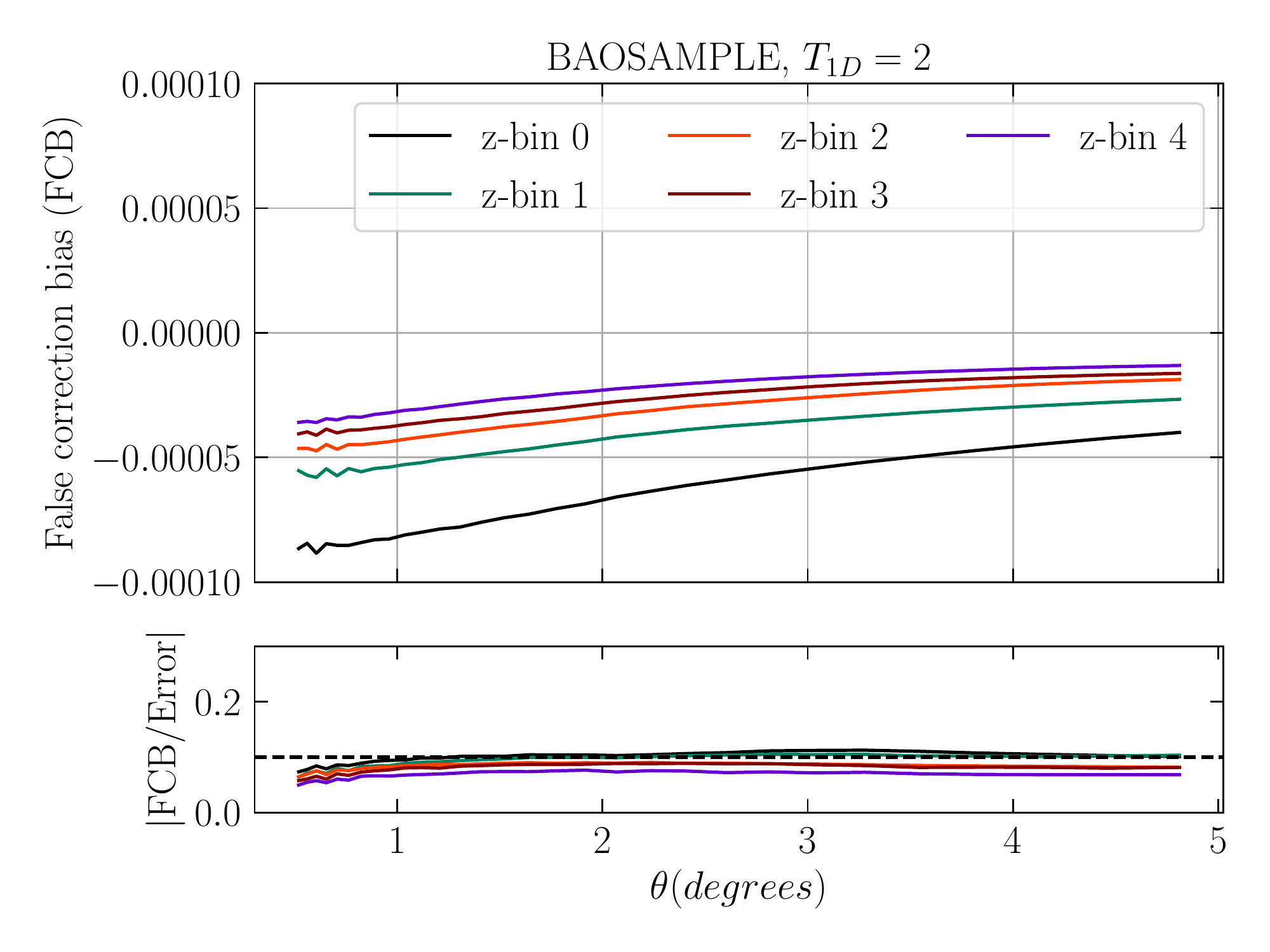}
\caption{False correction bias for $T_{1D}=4$ (top panel) and $T_{1D}=2$ (bottom panel). It measures the level of bias in $\omega(\theta)$ that might be introduced when we mistake statistical variations in the mocks by a true correlation with the SP maps. We also show the amplitude of the bias with respect to the statistical error. The comparison between different thresholds is one of the arguments used to define $T_{1D}=4$ (equivalent to a confidence level $\geq 99\%$) for the \baosample. The dashed horizontal line marks the 10\% level.}
    \label{fig:fcb_2}
\end{figure}

\begin{figure}
    \includegraphics[width=\linewidth]{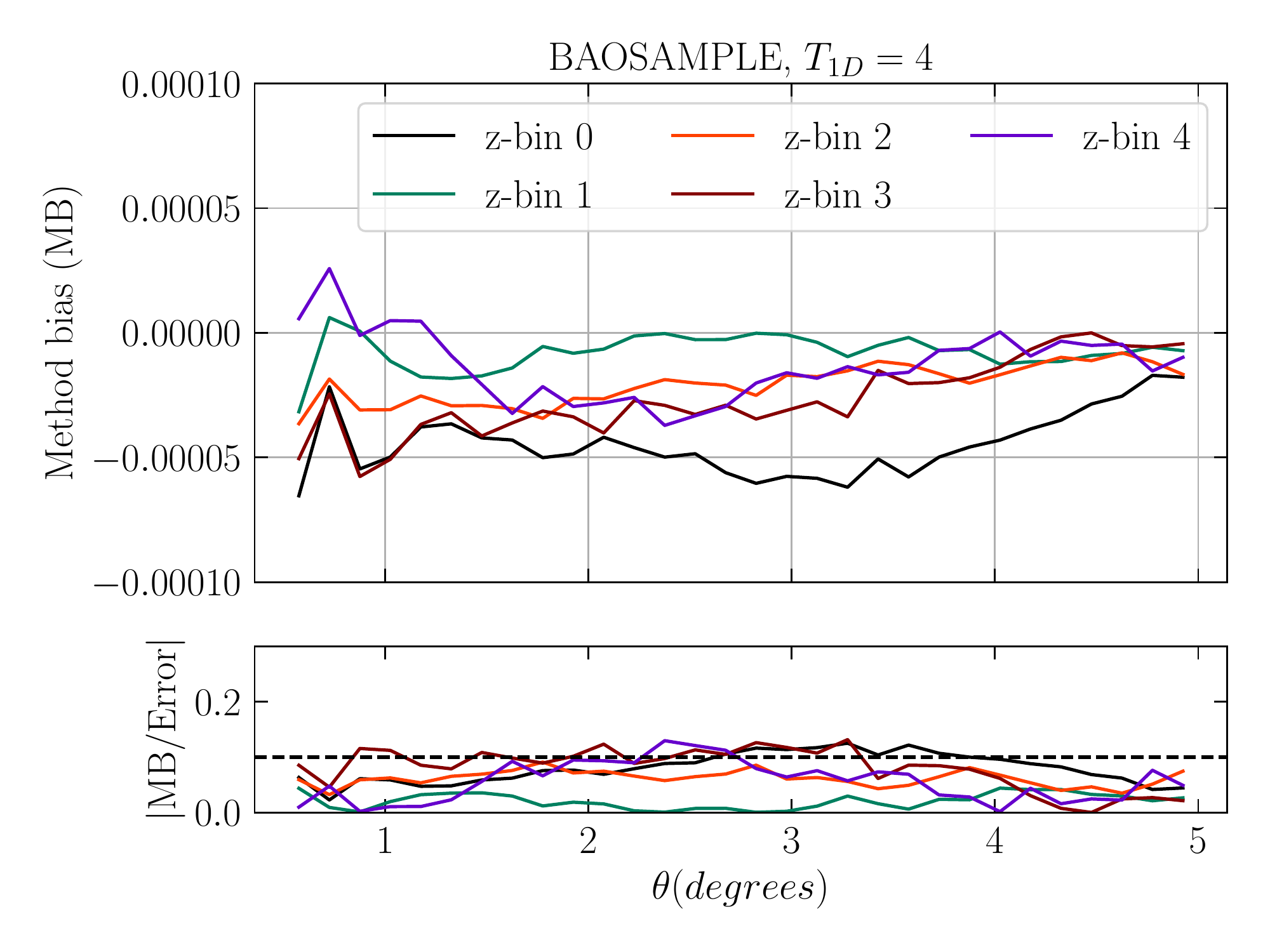}
\caption{Method bias results, which measure the bias we introduce in $\omega(\theta)$ after correcting for the weights, compared to the ``true'' $\omega(\theta)$. Bottom plot is the amplitude of the bias with respect to the statistical error. The dashed horizontal line marks the 10\% level.}
    \label{fig:eb_4}
\end{figure}

\begin{figure}
    \includegraphics[width=\linewidth]{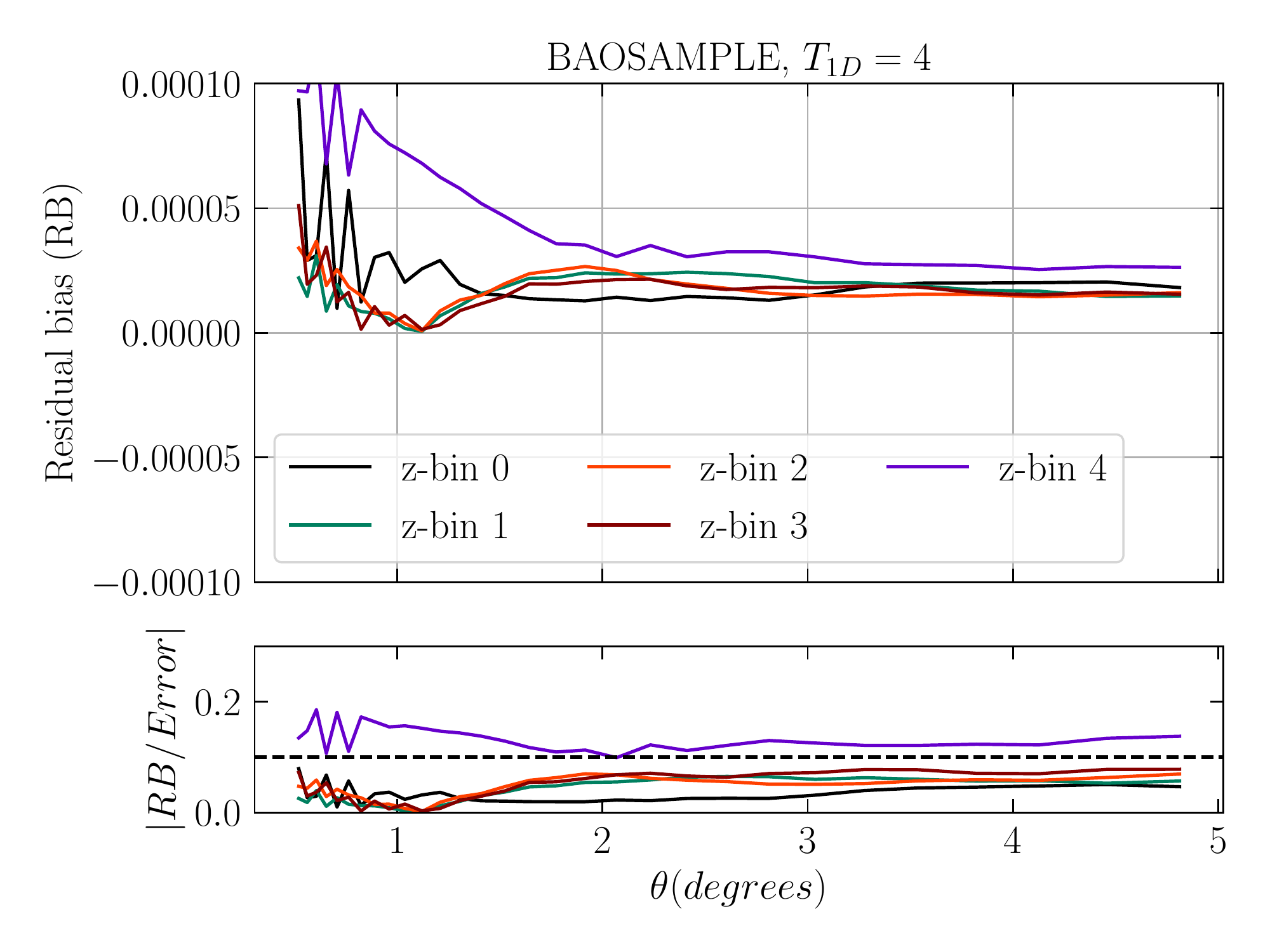}
\caption{Residual bias for $T_{1D}=4$. In the method bias, we fixed the list of SP maps used to run the method on the contaminated mocks, in the residual bias, we apply the same method, but this time, without choosing a pre-defined list of SP maps to correct for. Bottom plot is the amplitude of the bias with respect to the statistical error. The dashed horizontal line marks the 10\% level.}
    \label{fig:rsb_4}
\end{figure}

These validation tests are run over the mock realizations (the log-normal mocks presented in \secref{lognormalmocks}). A negative bias value will mean that we are ``over-correcting'' $\omega(\theta)$, a positive value that we are not correcting completely. Here we summarize our findings:

\begin{itemize}

 \item False correction bias: this effect measures the level of bias in $\omega(\theta)$ that we introduce when a chance correlation of a given realization of a SP map correlates with the cosmological structure. For this reason, we run the ISD method on uncontaminated mocks to evaluate how often this occurs, and it depends on the threshold used. A very strict threshold might imply that we will end correcting a spurious correlation. In \figref{fcb_2} we show the result for $T_{1D}=4$ and $T_{1D}=2$. In the case of $T_{1D}=2$, we find a bias that for some cases is $\roughly 10\%$ the statistical error, while for the case of $T_{1D}=4$, the false correction bias is negligible, justifying the choice of $T_{1D}=4$. These tests are done by applying the ISD method over the ``true'' un-contaminated mocks.
 
    \item Method bias: this effect measures the bias we introduce in $\omega(\theta)$ after correcting for the SP maps defined in the \baosample, compared to the ``true'' $\omega(\theta)$. It starts by contaminating the mocks with the same weight found in the data, and then apply the ISD method in each mock, recovering a de-contaminated set of mocks. It is defined as the mean difference between the de-contaminated and the ``true'' $\omega(\theta)$ from all mocks. We find the method bias to be $\lesssim 10\%$ the statistical error at all scales and redshifts (cf. \figref{eb_4}). 
    
       \item Residual bias: in the method bias, we fixed the list of SP maps used to those found in the data. In the residual bias test, we apply the same method, but this time, without fixing the list of SP maps, i.e.: each mock is free and independent with respect to the SP maps needed. As seen in \figref{rsb_4}, the bias amplitude reaches $\roughly 20\%$ the statistical error in the highest redshift bin, and it is below 10\% for the others.
\end{itemize}


The low level of bias found in all these tests validate the goodness of the method and, even if the correction level is high (as seen in \figref{deltatest_1}) we have demonstrated that the error introduced by the method is always below the statistical one. We conclude that the high correction introduced by observing conditions is due simply to the wide area of the survey, but that there are no pathological issues in the sample.   

\section{Unblinding $\omega(\theta)$}
\label{sec:unblinding}

At this point in the analysis, once we validated the systematic weights and characterized the \pz distributions, we freeze the \baosample and prepare for the unblinding of the 2-pt clustering measurements. To do so, we require the \baosample to pass a series of robustness tests (see~\citealt{y3mainbao}). If all tests are passed, we are ready to unblind the sample and measure the BAO scale. The sample presented in this paper passed all the robustness tests and therefore, it is the base for the BAO distance measurement in DES Y3. 

\begin{figure*}
    \includegraphics[width=\linewidth]{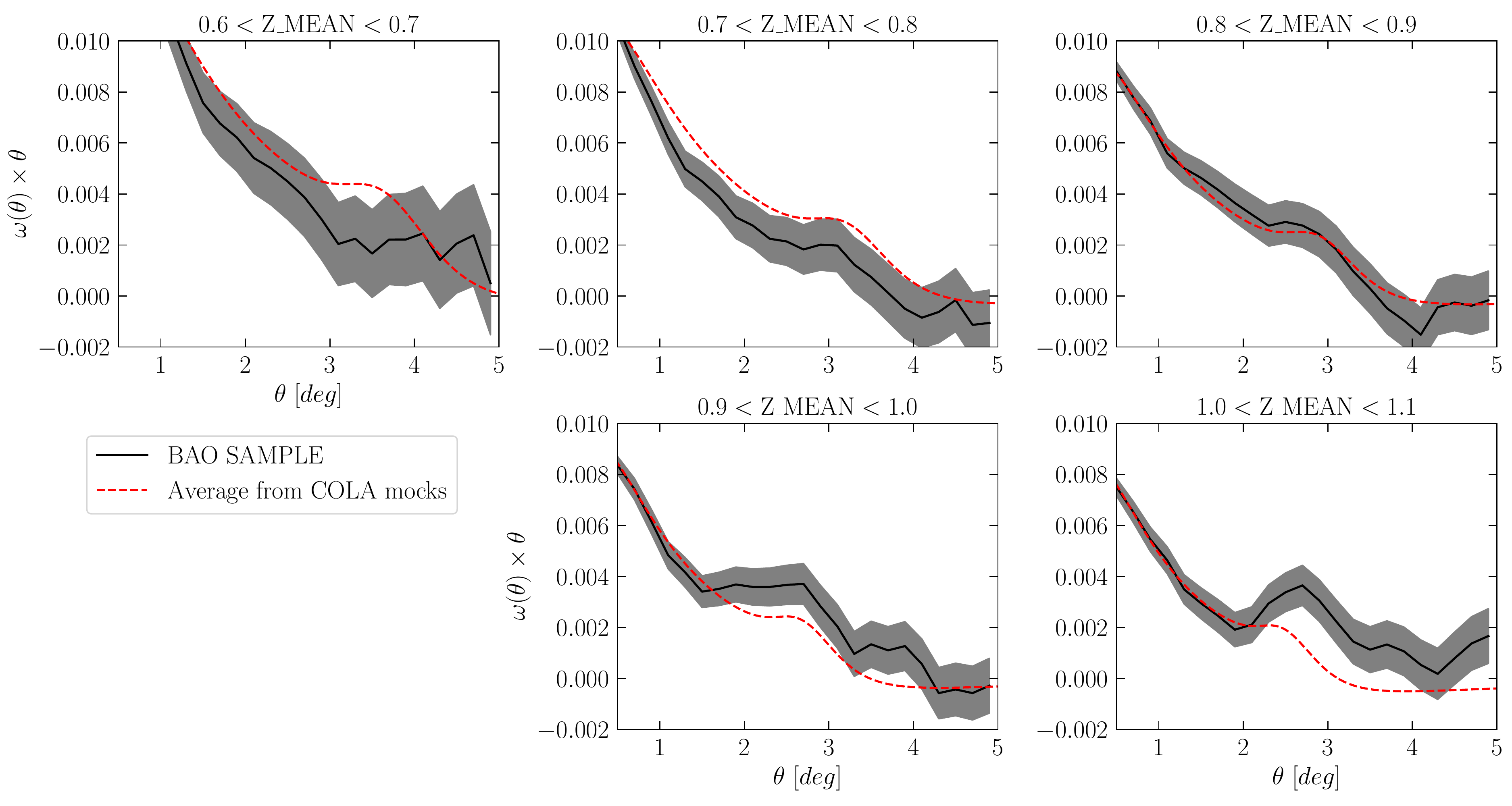}
\caption{$w(\theta)$ for the 5 tomographic redshift bins (shaded region). Errors are given by the diagonal elements of the covariance matrix coming from the COLA mocks~\citep{colamocks}, whose average value is shown in dashed red line. Cosmological implications from the position of the BAO peak are detailed in~\citet{y3mainbao}.}
    \label{fig:acf_data}
\end{figure*}

We present the angular correlation function $\omega(\theta)$, split into five redshift bins in~\figref{acf_data}. These measurements have been obtained using the standard Landy-Szalay estimator~\citep{landyszalay} with \var{CUTE}\footnote{\url{https://github.com/damonge/CUTE}}~\citep{cute}. Errors in these figures are estimated using the COLA mocks~\citep{colamocks}.

In~\citet{y3mainbao}, the \baosample is used to estimate the BAO scale at a mean effective redshift of 0.835, both in real and in configuration space, and we calculate the best-fit cosmology to it. Eventually, the BAO scale measurement will also be combined with other DES Y3 cosmological probes to obtain the most precise cosmological constraints from the combination of galaxy clustering and weak lensing. 

\section{Conclusions}
\label{sec:conclusions}

In this paper we describe the data used in DES Y3 for cosmological constraints from the BAO distance scale. Unlike Y1, and thanks to improvements of Y3 data, we have a larger density of galaxies that allows the extension of the analysis to redshift 1.1, increasing the effective redshift of the sample from 0.81 to 0.835. The sample covers 4100 square degrees to a depth of $i_{AB} \leq 22.3$ for galaxies with signal-to-noise greater than $10\sigma$. The galaxy selection is done based on the same colour scheme applied in Y1. This selection ensures a good \pz estimate, as attested by our \pz validation for galaxies with \pz $> 0.5$. Also, good quality \pz's ensure the \nz of the samples are well characterized. In the Y3 sample, we mitigate spurious clustering that arises from observing conditions with the same methodology used in the Y1 lensing samples. These weights are later applied to the \baosample in order to obtain un-biased angular correlation functions and power spectra. Cosmological implications of these results are later discussed in~\citet{y3mainbao}.

\section*{Acknowledgments}

We are grateful for the extraordinary contributions of our CTIO colleagues and the DECam Construction, Commissioning and Science Verification
teams in achieving the excellent instrument and telescope conditions that have made this work possible.  The success of this project also 
relies critically on the expertise and dedication of the DES Data Management group.

Funding for the DES Projects has been provided by the U.S. Department of Energy, the U.S. National Science Foundation, the Ministry of Science and Education of Spain, 
the Science and Technology Facilities Council of the United Kingdom, the Higher Education Funding Council for England, the National Center for Supercomputing 
Applications at the University of Illinois at Urbana-Champaign, the Kavli Institute of Cosmological Physics at the University of Chicago, 
the Center for Cosmology and Astro-Particle Physics at the Ohio State University,
the Mitchell Institute for Fundamental Physics and Astronomy at Texas A\&M University, Financiadora de Estudos e Projetos, 
Funda{\c c}{\~a}o Carlos Chagas Filho de Amparo {\`a} Pesquisa do Estado do Rio de Janeiro, Conselho Nacional de Desenvolvimento Cient{\'i}fico e Tecnol{\'o}gico and 
the Minist{\'e}rio da Ci{\^e}ncia, Tecnologia e Inova{\c c}{\~a}o, the Deutsche Forschungsgemeinschaft and the Collaborating Institutions in the Dark Energy Survey. 

The Collaborating Institutions are Argonne National Laboratory, the University of California at Santa Cruz, the University of Cambridge, Centro de Investigaciones Energ{\'e}ticas, 
Medioambientales y Tecnol{\'o}gicas-Madrid, the University of Chicago, University College London, the DES-Brazil Consortium, the University of Edinburgh, 
the Eidgen{\"o}ssische Technische Hochschule (ETH) Z{\"u}rich, 
Fermi National Accelerator Laboratory, the University of Illinois at Urbana-Champaign, the Institut de Ci{\`e}ncies de l'Espai (IEEC/CSIC), 
the Institut de F{\'i}sica d'Altes Energies, Lawrence Berkeley National Laboratory, the Ludwig-Maximilians Universit{\"a}t M{\"u}nchen and the associated Excellence Cluster Universe, 
the University of Michigan, NSF's NOIRLab, the University of Nottingham, The Ohio State University, the University of Pennsylvania, the University of Portsmouth, 
SLAC National Accelerator Laboratory, Stanford University, the University of Sussex, Texas A\&M University, and the OzDES Membership Consortium.

Based in part on observations at Cerro Tololo Inter-American Observatory at NSF's NOIRLab (NOIRLab Prop. ID 2012B-0001; PI: J. Frieman), which is managed by the Association of Universities for Research in Astronomy (AURA) under a cooperative agreement with the National Science Foundation.

The DES data management system is supported by the National Science Foundation under Grant Numbers AST-1138766 and AST-1536171.
The DES participants from Spanish institutions are partially supported by MICINN under grants ESP2017-89838, PGC2018-094773, PGC2018-102021, SEV-2016-0588, SEV-2016-0597, and MDM-2015-0509, some of which include ERDF funds from the European Union. IFAE is partially funded by the CERCA program of the Generalitat de Catalunya.
Research leading to these results has received funding from the European Research
Council under the European Union's Seventh Framework Program (FP7/2007-2013) including ERC grant agreements 240672, 291329, and 306478.
We  acknowledge support from the Brazilian Instituto Nacional de Ci\^encia
e Tecnologia (INCT) do e-Universo (CNPq grant 465376/2014-2).

This manuscript has been authored by Fermi Research Alliance, LLC under Contract No. DE-AC02-07CH11359 with the U.S. Department of Energy, Office of Science, Office of High Energy Physics.

This paper uses data from the VIMOS Public Extragalactic Redshift Survey (VIPERS). VIPERS has been performed using the ESO Very Large Telescope, under the ``Large Programme'' 182.A-0886. The participating institutions and funding agencies are listed at \url{http://vipers.inaf.it}.

A.C.R. acknowledges financial support from the Spanish Ministry of Science, Innovation and Universities (MICIU) under grant AYA2017-84061-P, co-financed by FEDER (European Regional Development Funds) and by the Spanish Space Research Program ``Participation in the NISP instrument and preparation for the science of EUCLID'' (ESP2017-84272-C2-1-R). S.A. was supported by the MICUES project, funded by the EU's H2020 MSCA grant agreement no. 713366 (InterTalentum UAM).



\section*{Data availability}
A general description of DES data releases is available on the survey website at \url{https://www.darkenergysurvey.org/the-des-project/data-access/}. \baosample is available on the DES Data Management website hosted by the National Center for Supercomputing Applications at \url{https://des.ncsa.illinois.edu/releases/y3a2}. 

\bibliographystyle{mnras}
\bibliography{biblio} 

\section{Affiliations}

$^{1}$ Instituto de Astrof\'isica de Canarias, E-38205 La Laguna, Tenerife, Spain \\
$^{2}$ Universidad de La Laguna, Dpto. Astrof\'isica, E-38206 La Laguna, Tenerife, Spain \\
$^{3}$ Laborat\'orio Interinstitucional de e-Astronomia - LIneA, Rua Gal. Jos\'e Cristino 77, Rio de Janeiro, RJ - 20921-400, Brazil\\
$^{4}$ Centro de Investigaciones Energ\'eticas, Medioambientales y Tecnol\'ogicas (CIEMAT), Madrid, Spain\\
$^{5}$ Institut d'Estudis Espacials de Catalunya (IEEC), 08034 Barcelona, Spain \\
$^{6}$ Institute of Space Sciences (ICE, CSIC),  Campus UAB, Carrer de Can Magrans, s/n,  08193 Barcelona, Spain \\
$^{7}$ Center for Cosmology and Astro-Particle Physics, The Ohio State University, Columbus, OH 43210, USA \\
$^{8}$ Department of Physics, The Ohio State University, Columbus, OH 43210, USA\\
$^{9}$ Institute of Theoretical Astrophysics, University of Oslo. P.O. Box 1029 Blindern, NO-0315 Oslo, Norway \\
$^{10}$ Physics Department, 2320 Chamberlin Hall, University of Wisconsin-Madison, 1150 University Avenue Madison, WI  53706-1390 \\ 
$^{11}$ Department of Physics, University of Michigan, Ann Arbor, MI 48109, USA \\
$^{12}$ Instituto de F\'isica Te\'orica, Universidade Estadual Paulista, S\~ao Paulo, Brazil \\
$^{13}$ Instituto de F\'isica Te\'orica UAM/CSIC, Universidad Aut\'onoma de Madrid, 28049 Madrid, Spain \\
$^{14}$ Departamento de F\'isica Te\'orica, Facultad de Ciencias, Universidad Aut\'onoma de Madrid, E-28049 Cantoblanco, Madrid, Spain \\
$^{15}$ Instituto de F\'isica Gleb Wataghin, Universidade Estadual de Campinas, 13083-859, Campinas, SP, Brazil \\
$^{16}$ Jet Propulsion Laboratory, California Institute of Technology, 4800 Oak Grove Dr., Pasadena, CA 91109, USA \\
$^{17}$ Kavli Institute for Particle Astrophysics \& Cosmology, P. O. Box 2450, Stanford University, Stanford, CA 94305, USA \\
$^{18}$ ICTP South American Institute for Fundamental Research, Instituto de F\'{\i}sica Te\'orica, Universidade Estadual Paulista, S\~ao Paulo, Brazil \\
$^{19}$ Department of Astronomy, University of Geneva, ch. d'\'Ecogia 16, CH-1290 Versoix, Switzerland \\
$^{20}$ Cerro Tololo Inter-American Observatory, NSF's National Optical-Infrared Astronomy Research Laboratory, Casilla 603, La Serena, Chile \\
$^{21}$ Fermi National Accelerator Laboratory, P. O. Box 500, Batavia, IL 60510, USA",0000-0002-7069-7857 \\
$^{22}$ CNRS, UMR 7095, Institut d'Astrophysique de Paris, F-75014, Paris, France \\
$^{23}$ Sorbonne Universit\'es, UPMC Univ Paris 06, UMR 7095, Institut d'Astrophysique de Paris, F-75014, Paris, France \\
$^{24}$ Department of Physics \& Astronomy, University College London, Gower Street, London, WC1E 6BT, UK \\
$^{25}$ Department of Astronomy and Astrophysics, University of Chicago, Chicago, IL 60637, USA \\
$^{26}$ SLAC National Accelerator Laboratory, Menlo Park, CA 94025, USA \\
$^{27}$ School of Mathematics and Physics, University of Queensland,  Brisbane, QLD 4072, Australia \\
$^{28}$ INAF, Astrophysical Observatory of Turin, I-10025 Pino Torinese, Italy \\
$^{29}$ Center for Astrophysical Surveys, National Center for Supercomputing Applications, 1205 West Clark St., Urbana, IL 61801, USA \\
$^{30}$ Department of Astronomy, University of Illinois at Urbana-Champaign, 1002 W. Green Street, Urbana, IL 61801, USA \\
$^{31}$ Institut de F\'{\i}sica d'Altes Energies (IFAE), The Barcelona Institute of Science and Technology, Campus UAB, 08193 Bellaterra (Barcelona) Spain \\
$^{32}$ Jodrell Bank Center for Astrophysics, School of Physics and Astronomy, University of Manchester, Oxford Road, Manchester, M13 9PL, UK \\
$^{33}$ University of Nottingham, School of Physics and Astronomy, Nottingham NG7 2RD, UK \\
$^{34}$ Astronomy Unit, Department of Physics, University of Trieste, via Tiepolo 11, I-34131 Trieste, Italy \\
$^{35}$ INAF, Observatorio Astronomico di Trieste, via G. B. Tiepolo 11, I-34143 Trieste, Italy \\
$^{36}$ Institute for Fundamental Physics of the Universe, Via Beirut 2, 34014 Trieste, Italy \\
$^{37}$ Observat\'orio Nacional, Rua Gal. Jos\'e Cristino 77, Rio de Janeiro, RJ - 20921-400, Brazil \\
$^{38}$ School of Mathematics and Physics, University of Queensland,  Brisbane, QLD 4072, Australia \\
$^{39}$ Department of Physics, IIT Hyderabad, Kandi, Telangana 502285, India \\
$^{40}$ Kavli Institute for Cosmological Physics, University of Chicago, Chicago, IL 60637, USA \\
$^{41}$ Department of Physics and Astronomy, University of Pennsylvania, Philadelphia, PA 19104, USA \\
$^{42}$ Santa Cruz Institute for Particle Physics, Santa Cruz, CA 95064, USA \\
$^{43}$ Department of Astronomy, University of Michigan, Ann Arbor, MI 48109, USA \\
$^{44}$ Institute of Astronomy, University of Cambridge, Madingley Road, Cambridge CB3 0HA, UK \\
$^{45}$ Kavli Institute for Cosmology, University of Cambridge, Madingley Road, Cambridge CB3 0HA, UK \\
$^{46}$ Centre for Astrophysics \& Supercomputing, Swinburne University of Technology, Victoria 3122, Australia \\
$^{47}$ Faculty of Physics, Ludwig-Maximilians-Universit\""at, Scheinerstr. 1, 81679 Munich, Germany \\
$^{48}$ Center for Astrophysics $\vert$ Harvard \& Smithsonian, 60 Garden Street, Cambridge, MA 02138, USA \\
$^{49}$ Lawrence Berkeley National Laboratory, 1 Cyclotron Road, Berkeley, CA 94720, USA \\
$^{50}$ Department of Astronomy/Steward Observatory, University of Arizona, 933 North Cherry Avenue, Tucson, AZ 85721-0065, USA \\
$^{51}$ Australian Astronomical Optics, Macquarie University, North Ryde, NSW 2113, Australia \\
$^{52}$ Lowell Observatory, 1400 Mars Hill Rd, Flagstaff, AZ 86001, USA \\
$^{53}$ Sydney Institute for Astronomy, School of Physics, A28, The University of Sydney, NSW 2006, Australia \\
$^{54}$ Centre for Gravitational Astrophysics, College of Science, The Australian National University, ACT 2601, Australia \\
$^{55}$ The Research School of Astronomy and Astrophysics, Australian National University, ACT 2601, Australia \\
$^{56}$ Departamento de F\'isica Matem\'atica, Instituto de F\'isica, Universidade de S\~ao Paulo, CP 66318, S\~ao Paulo, SP, 05314-970, Brazil \\
$^{57}$ George P. and Cynthia Woods Mitchell Institute for Fundamental Physics and Astronomy, and Department of Physics and Astronomy, Texas A\&M University, College Station, TX 77843,  USA \\
$^{58}$ Instituci\'o Catalana de Recerca i Estudis Avan\c{c}ats, E-08010 Barcelona, Spain \\
$^{59}$ Max Planck Institute for Extraterrestrial Physics, Giessenbachstrasse, 85748 Garching, Germany \\
$^{60}$ Universit\'e Clermont Auvergne, CNRS/IN2P3, LPC, F-63000 Clermont-Ferrand, France \\
$^{61}$ Department of Physics and Astronomy, University of Waterloo, 200 University Ave W, Waterloo, ON N2L 3G1, Canada \\
$^{62}$ Perimeter Institute for Theoretical Physics, 31 Caroline St. North, Waterloo, ON N2L 2Y5, Canada \\
$^{63}$ Department of Astrophysical Sciences, Princeton University, Peyton Hall, Princeton, NJ 08544, USA \\
$^{64}$ Brookhaven National Laboratory, Bldg 510, Upton, NY 11973, USA \\
$^{65}$ School of Physics and Astronomy, University of Southampton,  Southampton, SO17 1BJ, UK \\
$^{66}$ Computer Science and Mathematics Division, Oak Ridge National Laboratory, Oak Ridge, TN 37831 \\
$^{67}$ Institute of Cosmology and Gravitation, University of Portsmouth, Portsmouth, PO1 3FX, UK \\
$^{68}$ Department of Physics, Stanford University, 382 Via Pueblo Mall, Stanford, CA 94305, USA \\
$^{69}$ McDonald Observatory, The University of Texas at Austin, Fort Davis, TX 79734 \\


\appendix

\numberwithin{figure}{section}
\numberwithin{table}{section}

\section{MICE photometric validation}
\label{app:mice_phot}

In \secref{nz_samplevar} we measured the standard deviation of our \nz's by comparing the \baosample-VIPERS distribution to that from several realizations of the MICE simulation. To do so, we must be certain that the photometric properties of the MICE simulation match those from VIPERS and the \baosample. 

\begin{figure*}
\centering
\includegraphics[width=.9\linewidth]{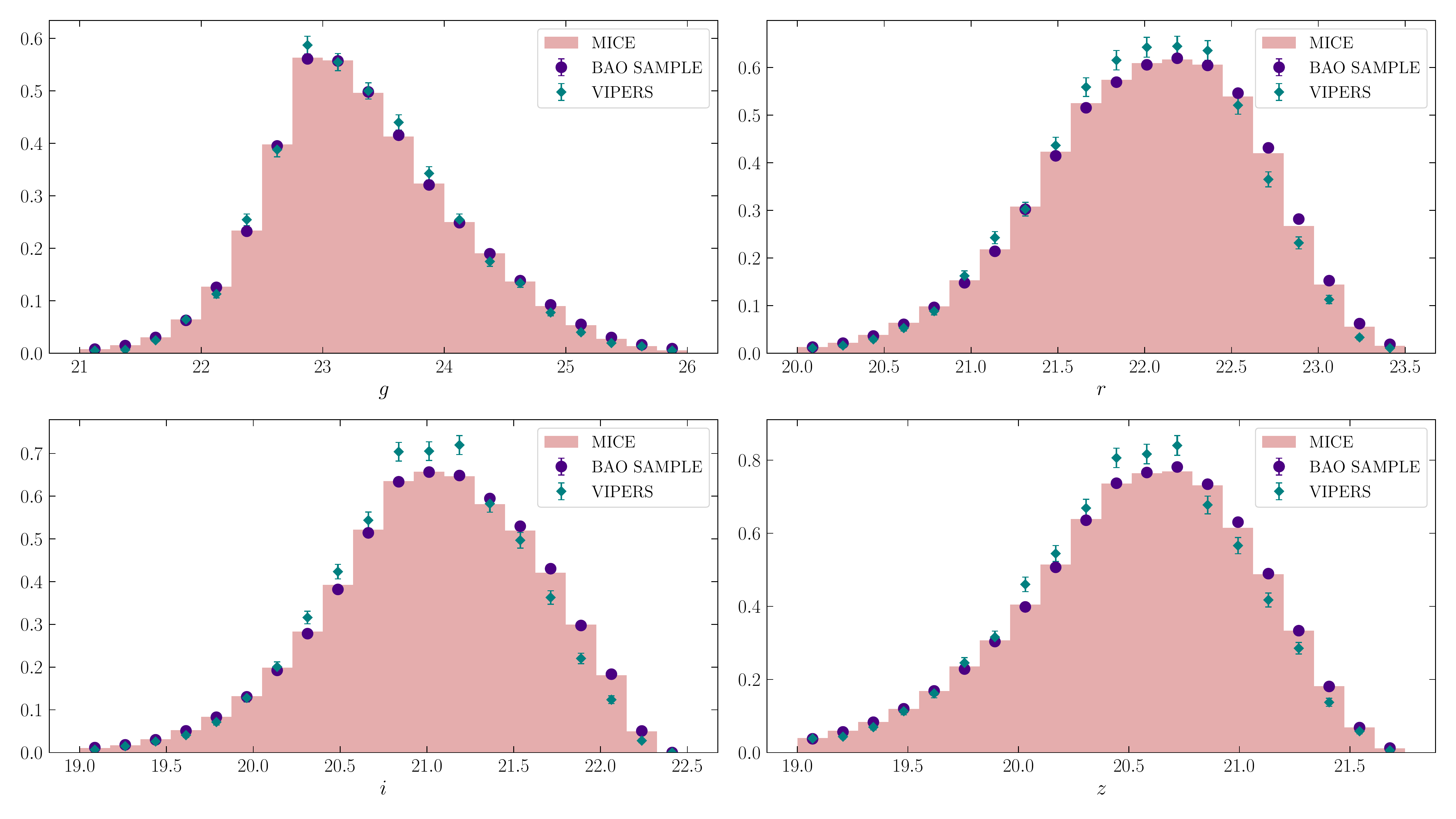}
\caption{Normalized distribution of the magnitudes in the different bands (top left: $g$, top right: $r$, bottom left: $i$, bottom right: $z$). Blue histogram: MICE data, orange points: DES data, green points: VIPERS data.}
\label{fig:hist_mag}
\end{figure*}

\begin{figure*}
\centering
\includegraphics[width=.9\linewidth]{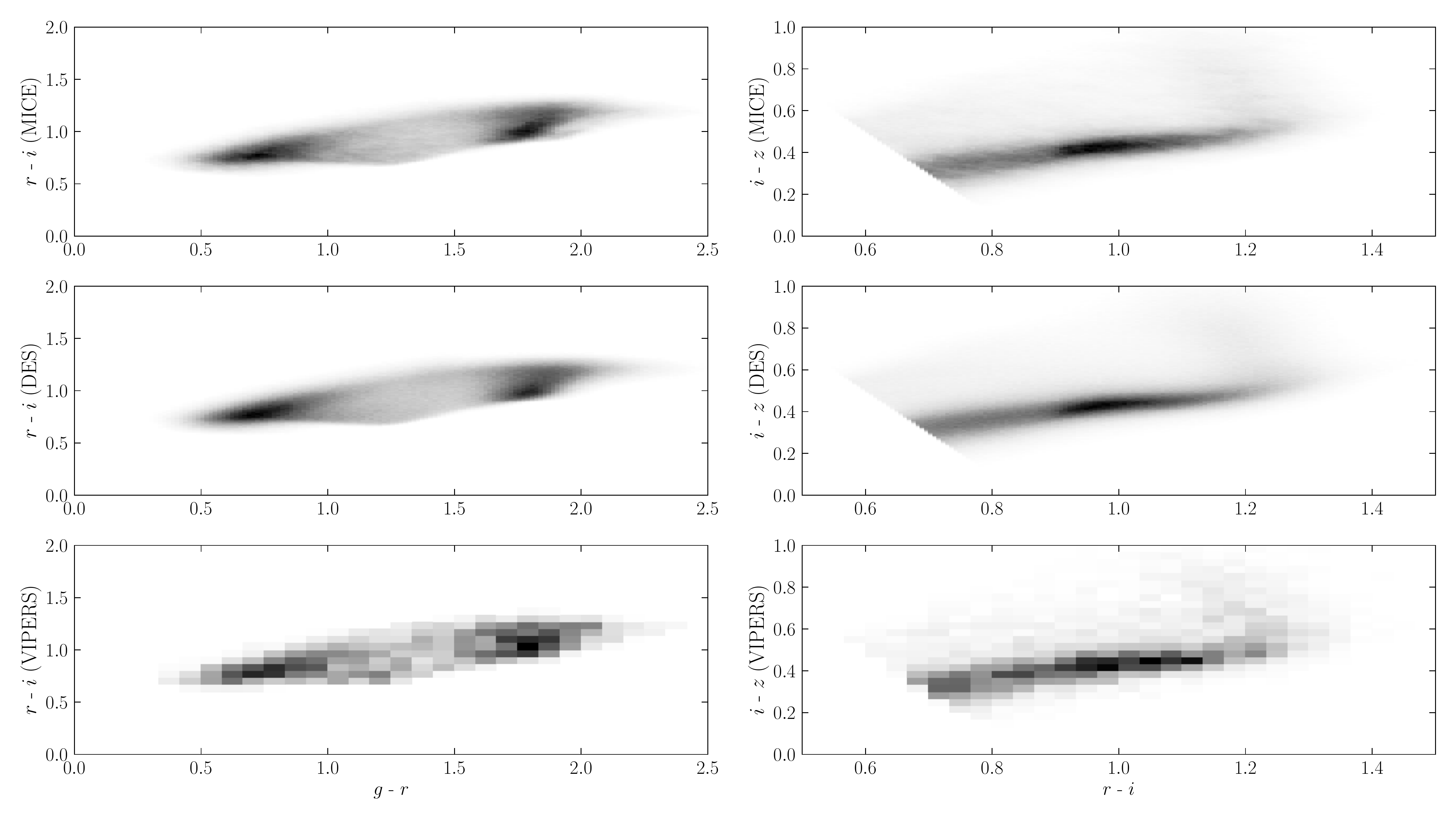}
\caption{Color-color distribution $g-r$ vs $r-i$ for MICE (top left panel), \baosample (middle left panel) and VIPERS (bottom left panel). Color-color distribution $r-i$ vs $i-z$ for MICE (top right panel), \baosample (middle right panel) and VIPERS (bottom right panel).}
\label{fig:colorcolor_diagrams}
\end{figure*}

In this Appendix we show the photometric comparisons between the \baosample and the MICE mocks. In \figref{hist_mag} we show the magnitude distribution of the samples in $griz$. In \figref{colorcolor_diagrams} we show the comparison in colour-colour space. 

The photometric properties of MICE, \baosample and VIPERS are very similar, hence, we can use this simulation to estimate errors in our \nz's. 

\section{Selection of representative SP maps}
\label{app:rep_spmaps}

\begin{figure*}
    \centering
    \includegraphics[width=.7\linewidth]{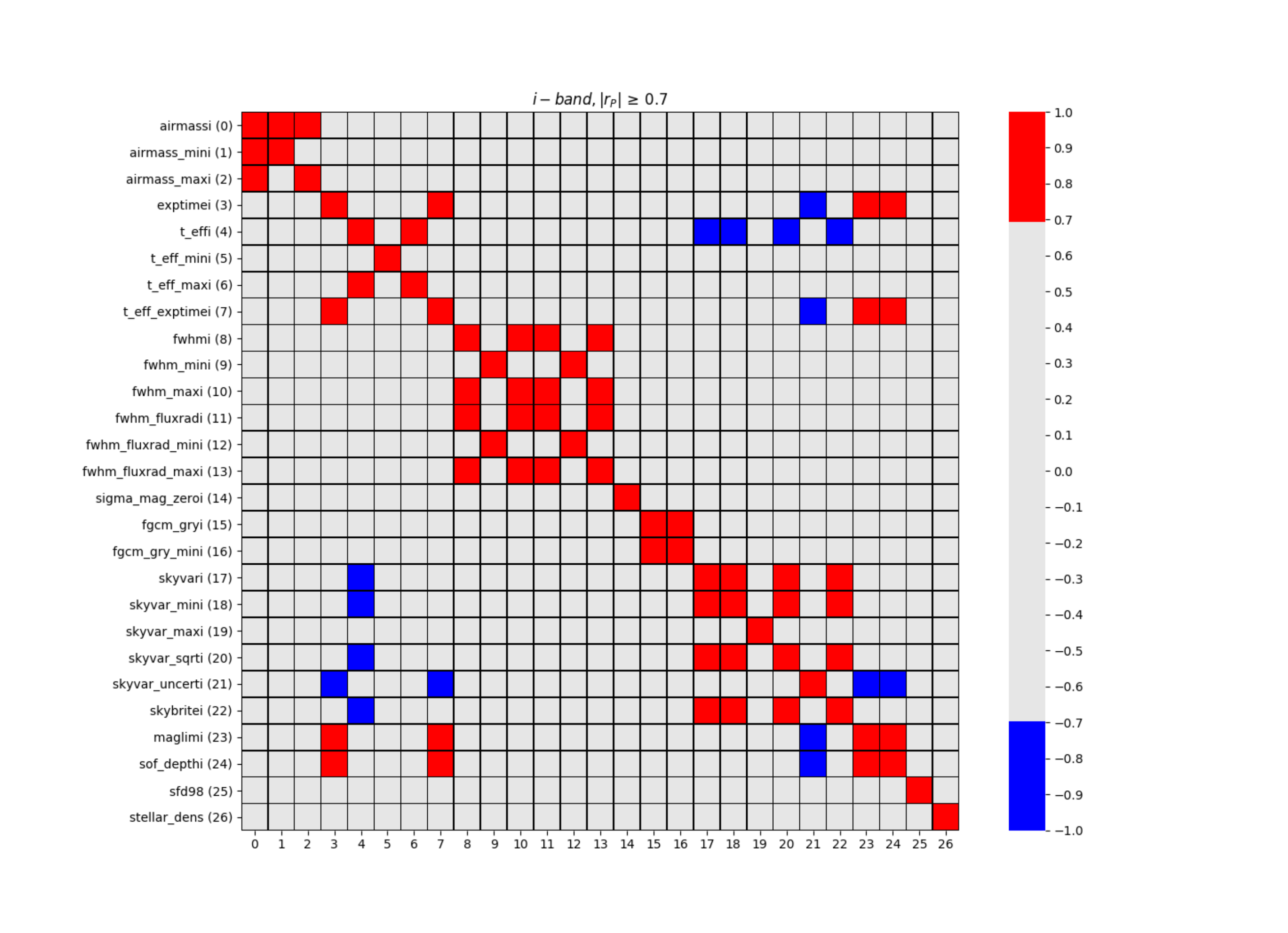}
    \caption{Pearson's correlation matrix for the $i$-band SP maps. Red (blue) cells correspond to the SP map pairs with correlation coefficient higher (lower) than $r_l = 0.7$. This helps us not only to identify SP map groups formed by the different statistics of the same observing condition, but also correlations among maps out of these groups, as for example the expected correlation between depth and exposure time. }
    \label{fig:matrix1}
\end{figure*}

As explained in \secref{lsssys}, ISD is an iterative process that evaluates the significance of each SP map at each redshift bin of the \baosample. Moreover, the 1D relations are calculated on a set of 1000 log-normal mocks, also at each redshift bin. This leads to a huge amount of computing time. For this reason, in order to optimize the iterative process that we applied to decontaminate the data, we reduce the number of SP maps that the pipeline has to run over. For this purpose, we look at their Pearson's correlation coefficients and set a lower limit $r_l$ for them. This way, we identify groups of highly correlated SP maps by looking at those maps with $\lvert r_P \rvert \geq r_l$ (cf. Figure \ref{fig:matrix1}). 

To check the stability of these groups with respect to the selection of $r_l$, we evaluate the correlation matrix ranging the value of $r_l$ from $0.5$ to $0.9$. Below $r_l = 0.5$ many SP maps are considered correlated (in the extreme case of a very low $r_l$ the whole matrix is considered a single group, so in that case an alternative would be to perform a PCA of the SP maps), while above $r_l = 0.9$ most of the maps are considered independent (there are almost no off-diagonal elements), not allowing us to reduce the number of maps, as desired. We observe a stable group structure between $r_l = 0.5$ and $0.7$, finally taking the highest value as our limit. Once we identify the SP map groups from the correlation matrix, we select a representative map from each of them. The representative SP maps chosen are listed in~\secref{spmaps}. 
     
We use the weighted average SP maps in all cases with the exception of the zero point residues, for which we use the total SP maps. To further ensure the stability of our results under the choice of these representative maps, we run ISD with slightly different lists of representative maps using different statistics within each SP map group. Since the number of maps needed to weight for is similar in all cases, we conclude that our results are stable under this choice. Furthermore, we apply this process for each photometric band separately, and we check that the same list of representative SP maps works for the four of them. 

\begin{figure*}
    \centering
    \includegraphics[width=.7\linewidth]{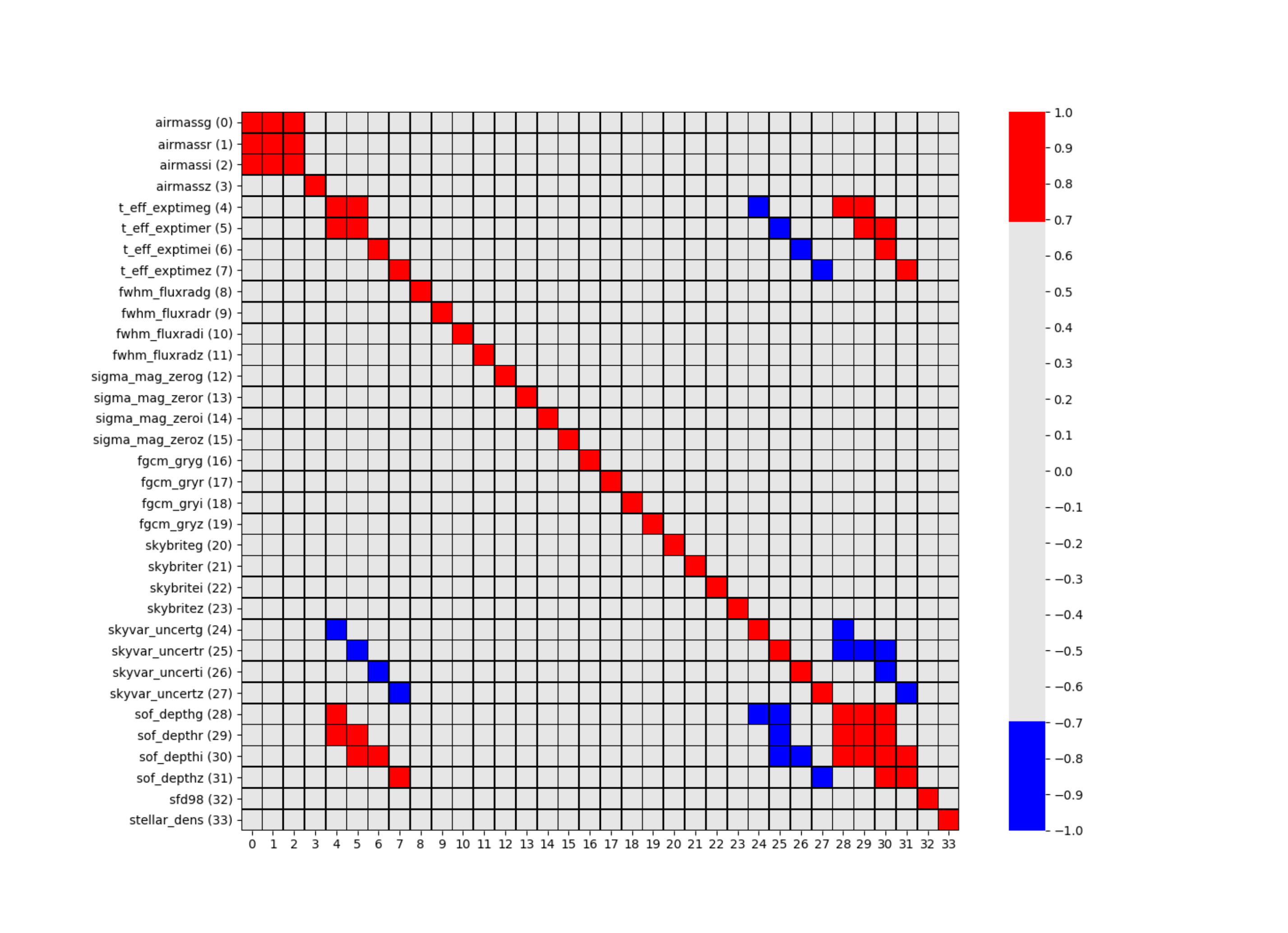}
    \caption{Correlation matrix for the representative SP maps in all photometric bands. The structures are similar equivalent to those seen in the case of individual photometric bands. }
    \label{fig:matrix3}
\end{figure*}

In \figref{matrix3} we show the correlation matrix for the final representative SP maps in $i$-band and the correlation of these sets of maps in the four bands that we work with, respectively. The remaining correlations at the same photometric band or among them are finally dealt by the correction pipeline. 

\section{Effect of weighting in BAO measurement}
\label{app:choicethreshold}

We run forecasts on BAO measurement using the log-normal mocks, applying a template-based method to recover the BAO scale~\citep{Seo_2012,2012MNRAS.427.2146X,2014MNRAS.441...24A,2017MNRAS.464.1168R}. This method estimates how different the BAO position is with respect to an assumed template cosmology. The difference is encoded in the parameter $\alpha$, which re-scales the separation between the BAO scale position in the theory and observation. 

\begin{figure}
        \centering
        \includegraphics[width=1.\linewidth]{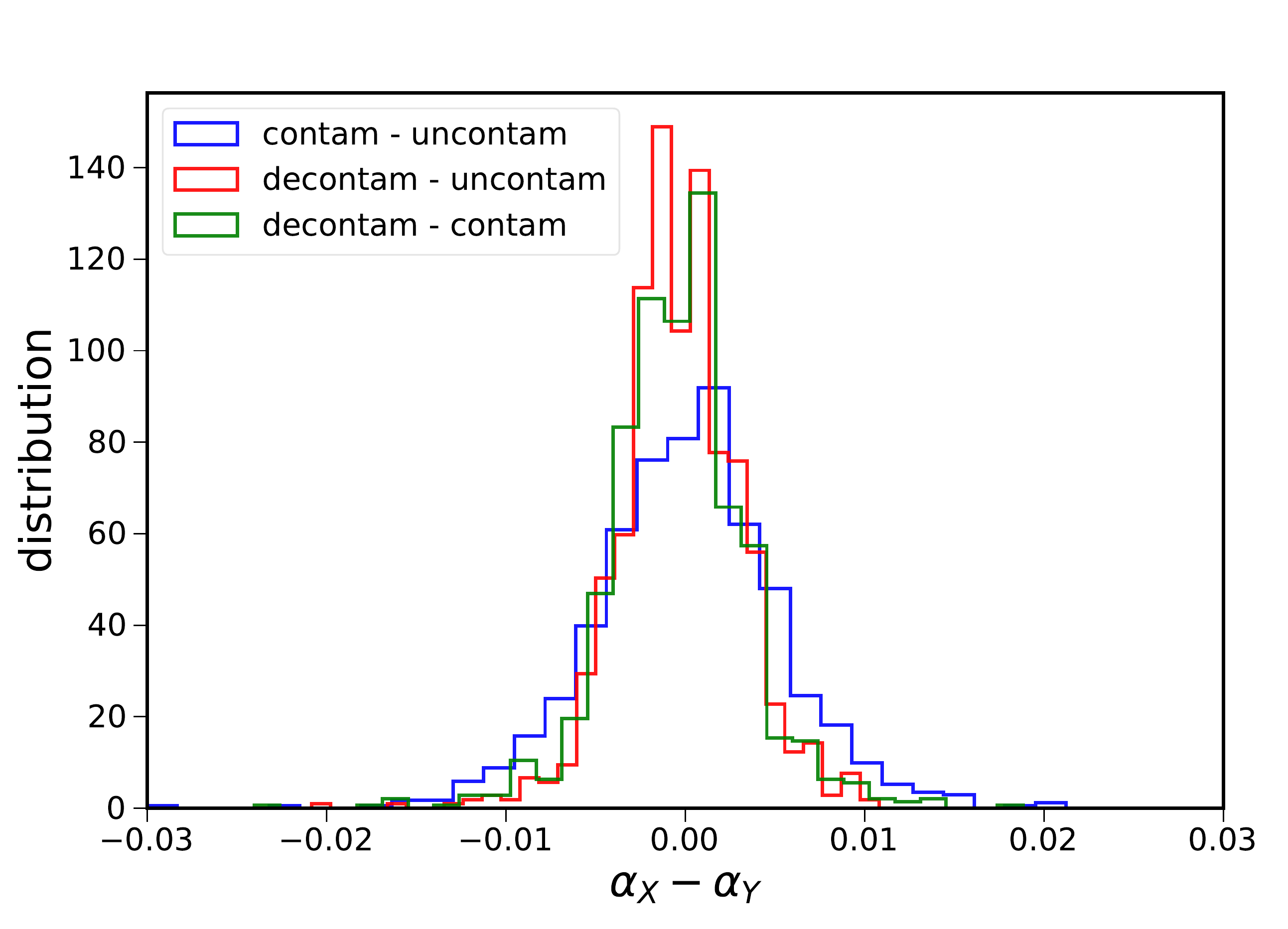}
        \caption{Difference in $\alpha$ found in three sets of mocks: \emph{uncontam} are uncontaminated mocks, with no systematic effects. \emph{contam} are the same mocks, contaminated with weights obtained at $T_{1D}=2$ and \emph{decontam} are the same contaminated mocks, after applying the ISD method at $T_{1D}=4$. The differences are compatible, confirming that the effect of weighting is small in BAO measurement.}
        \label{fig:alpha_diff}
    \end{figure}

We run it for uncontaminated mocks (without any systematic effect), in contaminated mocks with weights obtained at $T_{1D}=2$, and also, for de-contaminated mocks after applying the ISD scheme with $T_{1D}=4$ on the previous mocks, i.e., with a more relaxed threshold than the input contamination. The summary of this test can be found in \figref{alpha_diff}. Interestingly, we find the measurement works very well independent of whether we correct or not for observational systematics, as well as for the decontaminated mocks. In \tabref{alpha_diff} we show the recovered $\alpha$ in each case. This result is due to the fact that the contamination in the $\omega (\theta)$ is flat and does not affect the BAO scale. This finding is another argument to select the $T_{1D}=4$ in the analysis, instead of $T_{1D}=2$.

\begin{table*}
\centering
\caption{Mean $\alpha$ value and standard deviation from 1000 mocks. We recover the true cosmology in all cases and with the same precision. $\alpha$ encodes the BAO scale position difference between observations and theory.}
\label{tab:alpha_diff}
\begin{tabular}{c c c c }
 & uncontaminated & contaminated & decontaminated \\
\hline
$\bar{\alpha} \pm \ std$ & $1.01 \pm 0.02$ &  $1.01 \pm 0.02$ & $1.01 \pm 0.02$ \\
\end{tabular}
\end{table*}

\bsp	
\label{lastpage}
\end{document}